\documentclass[12pt,a4paper]{article}
\usepackage[T1]{fontenc}
\usepackage{lmodern}
\usepackage[a4paper,hmargin=23.5mm,top=24mm,textheight=230mm]{geometry}
\usepackage{amsmath,amssymb}
\usepackage{graphicx}
\usepackage{subfigure}
\usepackage{booktabs}
\usepackage{array}
\usepackage{placeins}
\usepackage{indentfirst}
\usepackage{cite}
\usepackage{microtype}
\usepackage[colorlinks=true,linkcolor=blue,urlcolor=blue,citecolor=black,
  hypertexnames=false,bookmarksopen=true,pdfstartview=FitV,pdfpagemode=UseNone,
  pdftitle={Thermodynamics and phase transitions of nonlinearly scalarized black holes in Einstein-phantom scalar-Gauss-Bonnet theory},
  pdfauthor={Ming Zhang, De-Cheng Zou, Yun Soo Myung, Bo Liu, Lina Zhang}]{hyperref}

\newcommand{\ud}{\mathop{}\!\mathrm{d}}

\begin{document}

\renewcommand{\thefootnote}{\fnsymbol{footnote}}
\begin{center}
{{{\Large \bf Thermodynamics and phase transitions of nonlinearly scalarized black holes in Einstein-phantom scalar-Gauss-Bonnet theory}}}\\[8mm]

Ming Zhang$^{a}$\footnote{mingzhang0807@126.com}, De-Cheng Zou$^{b}$\footnote{Corresponding author: dczou@jxnu.edu.cn}, Yun Soo Myung$^{c}$\footnote{ysmyung@inje.ac.kr}, Bo Liu$^{d}$\footnote{fenxiao2001@163.com}, \\ You Long$^{a}$
\footnote{L191127208@163.com} and Lina Zhang$^{e}$\footnote{linazhang@hnit.edu.cn}\\[7mm]
${}^{a}$Faculty of Science, Xihang University, Xi'an 710077, China\\
${}^{b}$School of Physics, Jiangxi Normal University, Nanchang 330022, China\\
${}^{c}$Center for Quantum Spacetime, Sogang University, Seoul 04107, Republic of Korea\\
${}^{d}$School of Arts and Sciences, Shaanxi University of Science and Technology, Xi'an, 710021, China\\
${}^{e}$College of Science, Hunan Institute of Technology, Hengyang 421002, China
\end{center}
\renewcommand{\thefootnote}{\arabic{footnote}}
\setcounter{footnote}{0}

\begin{abstract}
In this paper, we investigate the thermodynamic properties of static,
nonlinearly scalarized black holes in Einstein--phantom scalar--Gauss--Bonnet
theory with polynomial and exponential couplings and compare them with
the corresponding canonical EsGB solutions.
The phantom lower branches have higher Wald entropy at fixed mass and
lower free energy at fixed temperature than Schwarzschild black holes.
The phantom exponential primary branches develop negative entropy at
small horizon radii. Positive heat capacity occurs between the mass and
temperature extrema. The phantom field reverses their order,
the direction of free-energy crossings, and the signs of the entropy and
mass jumps.
For both couplings, the phantom lower branches and portions of the primary
branches are therefore thermodynamically preferred over Schwarzschild
black holes.
\end{abstract}

\section{Introduction}
\label{intro}

Black holes with scalar hair have been widely studied in theories of
gravity beyond general relativity. However,
the no-hair theorem rules out stationary, asymptotically flat black holes
with a minimally coupled real scalar field under standard assumptions
\cite{Bekenstein:1995un,Herdeiro:2015waa}. This theorem can be
circumvented when the scalar field couples nonminimally to the
curvature. A well-studied example is the coupling to the Gauss--Bonnet
invariant, which allows black holes to carry scalar hair while keeping
the field equations of second order. Kanti et al. obtained dilatonic
black holes in higher-curvature string gravity \cite{Kanti:1995vq}, and
Sotiriou and Zhou found black hole hair in generalized scalar--tensor
gravity \cite{Sotiriou:2013qea}. Antoniou et al. later constructed hairy
black holes for more general Einstein--scalar--Gauss--Bonnet (EsGB)
couplings \cite{Antoniou:2017acq,Antoniou:2017hxj}.

In scalar--tensor theories, Damour and Esposito-Far\`ese found a
mechanism of spontaneous scalarization when studying neutron stars
\cite{Damour:1993hw}. Doneva and Yazadjiev, and Silva et al., showed
that black holes in EsGB theory can also undergo spontaneous
scalarization \cite{Doneva:2017bvd,Silva:2017uqg}. In a region of sufficiently strong
curvature, the Gauss--Bonnet coupling gives the scalar perturbation a
tachyonic effective mass. The Schwarzschild black hole then becomes
unstable at the linear level, and scalarized branches bifurcate from it
at the onset of the instability. The stability of these branches
depends on the higher-order terms in the coupling function
\cite{Blazquez-Salcedo:2018jnn,Silva:2018qhn} and on scalar
self-interactions \cite{Macedo:2019sem}. Radial and angular
perturbations impose further restrictions \cite{Minamitsuji:2024twp}.
A review of these developments can be found in
Ref.~\cite{Doneva:2022ewd}.
Couplings that begin at quartic order, however, allow nonlinear scalarization.
They satisfy $\zeta''(0)=0$, so scalar perturbations on the Schwarzschild
background are massless at the linear level. A perturbation of
sufficiently large amplitude can produce scalar hair. Doneva and
Yazadjiev found this nonlinear scalarization for an exponential coupling
\cite{Doneva:2021tvn}. It was then studied from several aspects,
including radial stability \cite{Blazquez-Salcedo:2022omw}, the effects
of scalar mass and self-interactions \cite{Pombo:2023lxg}, and rotating
black holes \cite{Doneva:2022yqu,Lai:2023gwe,Lai:2026vxs}. For
polynomial couplings, Zou et al. identified the threshold amplitudes
from the time evolution of scalar perturbations on the Schwarzschild
background, and constructed the fully backreacted static scalarized
black holes \cite{Zou:2024paper}.

Moreover, the thermodynamic properties of scalarized black
holes have also received increasing attention. In EsGB theory, the
entropy of a black hole contains the Bekenstein--Hawking area term
\cite{Bekenstein:1973ur,Hawking:1974sw} together with a Gauss--Bonnet
contribution given by the Wald formula \cite{Wald:1993nt,Iyer:1994ys}.
The entropy comparison with a Schwarzschild black hole of the same mass
depends on both contributions. At fixed temperature, the two black holes are
compared through the free energy obtained from the Euclidean action
\cite{Gibbons:1976ue}. The thermodynamics of canonical nonlinearly
scalarized black holes with polynomial and exponential couplings has
been studied in Refs.~\cite{Zou:2026thermo,Herdeiro:2026sur}, including
entropy and free-energy comparisons with Schwarzschild black holes and
a numerical check of the first law. The interpretation of the crossing
points depends on the choice of ensemble, as discussed in
Ref.~\cite{Herdeiro:2026sur}. An asymptotically flat
Schwarzschild black hole has negative heat capacity \cite{Gross:1982cv},
and placing the black hole in a finite cavity changes its thermodynamic
stability \cite{York:1986it}.

In the present work, we investigate the thermodynamic properties of
nonlinearly scalarized black holes in Einstein--phantom scalar--Gauss--Bonnet
(EPsGB) theory with polynomial and exponential couplings. The canonical solutions
with the same couplings allow us to assess the effects of the phantom
field on entropy and free-energy ordering. We verify the first law
numerically and determine the intervals with positive heat capacity near the
turning points. We also compare the canonical polynomial and exponential
couplings matched through order $\phi^8$ to examine their behavior at
finite scalar amplitude. We work with classical,
asymptotically flat solutions and use the heat capacity to assess local
thermal stability at the free-energy crossings.

This paper is organized as follows. Section~\ref{sec:numerical-solutions}
introduces the canonical and phantom theories, discusses nonlinear
scalarization on the Schwarzschild background, and constructs the static
scalarized black holes. Section~\ref{sec:thermodynamic-results} examines
temperature and entropy, verifies the first law, and then analyzes heat
capacity and free energy. The effects of the coupling function and the
phantom field are discussed alongside the corresponding thermodynamic
results. Section~\ref{sec:comparison} contains the discussion and
conclusions.

\section{Nonlinear scalarization and static solutions}
\label{sec:numerical-solutions}
\label{model}

In this section, we introduce the EPsGB model with polynomial and
exponential couplings. We then construct static scalarized black holes
and compare their properties with those of the corresponding canonical
EsGB solutions.

\subsection{Field equations and nonlinear scalarization}
\label{sec:common-theory}
\label{sec:probe-evolution}

The four-dimensional Einstein--scalar--Gauss--Bonnet action takes the form
\begin{equation}
S=\frac{1}{16\pi}\int \ud^4x\sqrt{-g}
\left[R+\epsilon\left(-2\nabla_\mu\phi\nabla^\mu\phi
+\lambda^2\zeta(\phi){\cal R}_{\rm GB}^2\right)\right],
\label{action}
\end{equation}
where $\lambda$ is the Gauss--Bonnet coupling constant and $\zeta(\phi)$
is the coupling function. The Gauss--Bonnet invariant is defined as
\begin{equation}
{\cal R}_{\rm GB}^2=R_{\mu\nu\rho\sigma}R^{\mu\nu\rho\sigma}
-4R_{\mu\nu}R^{\mu\nu}+R^2.
\label{gb}
\end{equation}

Varying the action \eqref{action} with respect to the metric and the
scalar field, we obtain the field equations
\begin{eqnarray}
&&G_{\mu\nu}+\epsilon\Gamma_{\mu\nu}
=\epsilon\left(2\nabla_\mu\phi\nabla_\nu\phi
-g_{\mu\nu}\nabla_\alpha\phi\nabla^\alpha\phi\right),
\label{Einstein}\\
&&\Box\phi+\frac{\lambda^2}{4}\zeta'(\phi){\cal R}_{\rm GB}^2=0,
\label{KG}
\end{eqnarray}
where the Gauss--Bonnet contribution to the metric equation is
\begin{align}
\Gamma_{\mu\nu}=4\lambda^2\bigg[&
 R_{\mu\rho\nu\sigma}\nabla^\rho\nabla^\sigma\zeta
 +2R_{\rho(\mu}\nabla_{\nu)}\nabla^\rho\zeta
 -g_{\mu\nu}R_{\rho\sigma}\nabla^\rho\nabla^\sigma\zeta\nonumber\\
 &-R_{\mu\nu}\Box\zeta
 +\frac R2\left(g_{\mu\nu}\Box\zeta-\nabla_\mu\nabla_\nu\zeta\right)\bigg],
\label{Gamma}
\end{align}
with $\zeta=\zeta(\phi)$, and the parentheses denote symmetrization.

To examine the role of the coupling function, we consider polynomial and
exponential forms in canonical EsGB and phantom EPsGB theory. The
polynomial model is specified by
\begin{equation}
\zeta_p(\phi)=\alpha\phi^4-\beta_p\phi^8,
\label{couplings}
\end{equation}
where we fix $\alpha=1/4$ and take $\beta_p=125$. We also include the
pure quartic limit, $\beta_p=0$, to assess the effect of the eighth-order
term. The exponential model takes the form
\begin{equation}
\zeta_e(\phi)=\frac{1-e^{-\beta_e\phi^4}}{4\beta_e},\qquad
\beta_e\in\{25,100,1000\}.
\label{eq:exp-coupling}
\end{equation}

For both couplings, the choice $\epsilon=+1$ gives canonical EsGB theory
\cite{Doneva:2017bvd,Silva:2017uqg,Antoniou:2017acq}, while
$\epsilon=-1$ reverses both the scalar kinetic and Gauss--Bonnet terms,
defining EPsGB theory. Phantom scalarization with a different coupling
was studied in Ref.~\cite{Li:2025vcq}.
The thermodynamic phase structure of canonical EsGB black holes has
already been analyzed for polynomial~\cite{Zou:2026thermo} and
exponential~\cite{Herdeiro:2026sur} couplings. We compare their EPsGB
counterparts with these solutions to determine how the phantom field
modifies entropy and free-energy ordering.

The scalar source in Eq.~\eqref{KG} is determined by the derivative of
the coupling function. For the exponential model, the relevant
derivatives and small-field expansion read
\begin{align}
\zeta'_e(\phi)&=\phi^3e^{-\beta_e\phi^4},&
\zeta''_e(\phi)&=\phi^2(3-4\beta_e\phi^4)e^{-\beta_e\phi^4},
\label{eq:exp-derivatives}\\
\zeta_e(\phi)&=\frac{\phi^4}{4}-\frac{\beta_e\phi^8}{8}
 +\frac{\beta_e^2\phi^{12}}{24}+O(\phi^{16}).
\label{eq:exp-series}
\end{align}
The two couplings agree through order $\phi^8$ when $\beta_e=8\beta_p$;
we use $(\beta_p,\beta_e)=(125,1000)$ for this comparison. Their nonlinear
sources nevertheless differ. For $\beta_p>0$, the polynomial derivative
changes sign at
\begin{equation}
\phi_*=(8\beta_p)^{-1/4},
\label{roots}
\end{equation}
while the derivatives of the pure quartic and exponential couplings
vanish only at $\phi=0$.

Both couplings satisfy $\zeta(0)=\zeta'(0)=\zeta''(0)=0$, so the
Schwarzschild solution with $\phi=0$ exists in canonical EsGB and phantom
EPsGB theory. Its linear scalar perturbations obey a massless wave
equation and exhibit no tachyonic instability. Scalar hair can instead be generated by
finite-amplitude perturbations
\cite{Doneva:2021tvn,Blazquez-Salcedo:2022omw,Pombo:2023lxg,Zou:2024paper}.
To illustrate this nonlinear mechanism in the probe limit, we consider
spherically symmetric scalar evolution on the Schwarzschild background.
The metric function, rescaled field and tortoise coordinate are defined by
\begin{eqnarray}
&&f(r)=1-\frac{2M}{r},\qquad \Psi=r\phi,\qquad
\frac{\ud r_*}{\ud r}=f^{-1}.
\label{bg:probetransform}
\end{eqnarray}
In these variables, the scalar equation \eqref{KG} becomes
\begin{eqnarray}
&&-\partial_t^2\Psi+\partial_{r_*}^2\Psi-\frac{2Mf}{r^3}\Psi
+\frac{12\lambda^2M^2f}{r^5}\zeta'\left(\frac{\Psi}{r}\right)=0.
\label{bg:wave}
\end{eqnarray}

The parameter $\epsilon$ cancels from the scalar equation because it
multiplies the entire scalar sector in the action \eqref{action}.
Equation~\eqref{bg:wave} is therefore identical in canonical EsGB and
phantom EPsGB theory, whose differences enter through the metric
backreaction in Eq.~\eqref{Einstein}.
We adopt the fixed-background approach of Ref.~\cite{Zou:2024paper} and
evolve initially static Gaussian pulses,
$\phi(0,r_*)=A\exp[-(r_*-r_{*c})^2/(2\sigma^2)]$, with amplitude $A$,
center $r_{*c}=-84.82$ and width $\sigma=1$. The numerical evolution uses
$r_*\in[-120,200]$, $\Delta r_*=0.125$ and $\Delta t=0.1$, with
fourth-order Runge--Kutta time integration and sixth-order spatial
finite differences. Ingoing and outgoing wave
conditions are imposed at the inner and outer boundaries, respectively.

For $M=1$ and $\lambda=40$, small-amplitude pulses decay. Larger pulses
reproduce the polynomial behavior reported in Ref.~\cite{Zou:2024paper}:
saturation near $|\phi_H|=\phi_*$ for $\beta_p=125$ and runaway growth in
the pure quartic limit. For the exponential coupling with $\beta_e=1000$,
the field approaches a plateau near $|\phi_H|=0.27315$.
The effective-potential interpretation of Ref.~\cite{Zou:2024paper}
relates these behaviors to
\begin{eqnarray}
&&V_{\rm eff}(r,\phi)=-\frac{12\lambda^2M^2}{r^6}\zeta(\phi),\qquad
\lim_{|\phi|\to\infty}V_{\rm eff}^{(e)}=-\frac{3\lambda^2M^2}{\beta_e r^6}.
\label{bg:localpotential}
\end{eqnarray}
For $\beta_p>0$, the minima at $\phi=\pm\phi_*$ account for polynomial
saturation. The pure quartic potential is unbounded from below.
The exponential potential instead approaches a finite negative limit,
without a minimum at finite positive $\phi$. The late-time scalar
plateau also depends on radial gradients and absorption at the horizon.

\subsection{Static scalarized black hole solutions}
\label{sec:common-construction}
\label{sec:backreacted-branches}
\label{sec:probe-limit}
\label{sec:probe-phantom-branches}
\label{sec:poly-branches}
\label{sec:poly-solutions}
\label{sec:exp-coverage}
\label{sec:exp-solutions}
\label{sec:phantom-geometry}

With scalar backreaction included, the metric of a static scalarized black
hole takes the spherically symmetric form
\begin{equation}
\ud s^2=-A(r)\ud t^2+\frac{\ud r^2}{B(r)}+r^2\ud\Omega_2^2.
\label{metric}
\end{equation}
Substituting this ansatz into Eqs.~\eqref{Einstein} and \eqref{KG},
we obtain the $tt$, $rr$ and scalar equations,
\begin{align}
0={}&1-B-rB'-\epsilon Br^2\phi'^2
 +2\epsilon\lambda^2(3B-1)B'\zeta'\phi'\nonumber\\
 &+4\epsilon\lambda^2B(B-1)(\zeta'\phi''+\zeta''\phi'^2),
\label{eq:radial-Et}\\
0={}&B\left[r+2\epsilon\lambda^2(1-3B)\zeta'\phi'\right]A'
 +A\left[B(1-\epsilon r^2\phi'^2)-1\right],
\label{eq:radial-Er}\\
0={}&\phi''+\frac{\phi'}2\left(\frac{A'}A+\frac{B'}B+\frac4r\right)\nonumber\\
 &+\frac{\lambda^2\zeta'}{r^2}\left[
 (B-1)\left(\frac{A''}A-\frac{A'^2}{2A^2}\right)
 +\frac{(3B-1)A'B'}{2AB}\right].
\label{eq:radial-Ephi}
\end{align}
Here the primes on $A$, $B$ and $\phi$ denote derivatives with respect
to $r$, and $\zeta'$ and $\zeta''$ denote derivatives with respect to
$\phi$. The $\theta\theta$ component of Eq.~\eqref{Einstein} gives
\begin{align}
0={}&\left(\frac B2-\frac{2\epsilon\lambda^2B^2\zeta'\phi'}r\right)
 \left(\frac{A''}A-\frac{A'^2}{2A^2}\right)\nonumber\\
 &+\left(\frac{B'}4-\frac{3\epsilon\lambda^2BB'\zeta'\phi'}r\right)\frac{A'}A
 +\frac{BA'/A+B'}{2r}+\epsilon B\phi'^2\nonumber\\
 &-\frac{2\epsilon\lambda^2B^2A'}{rA}(\zeta'\phi''+\zeta''\phi'^2).
\label{eq:radial-Eomega}
\end{align}

To construct the solutions, we impose a regular, nonextremal event
horizon at $r=r_H$. The near-horizon expansions of the metric functions
and scalar field are
\begin{equation}
\begin{aligned}
A&=a_1(r-r_H)+O((r-r_H)^2),\qquad
B=b_1(r-r_H)+O((r-r_H)^2),
\\
\phi&=\phi_H+\phi'_H(r-r_H)+O((r-r_H)^2),
\end{aligned}
\label{horizon}
\end{equation}
where $\phi_H=\phi(r_H)$, $a_1=A'(r_H)$ and $b_1=B'(r_H)$.

Substituting the expansions \eqref{horizon} into the field
equations fixes the scalar slope $\phi'_H$. Choosing the root connected
continuously to the limit of vanishing coupling, we obtain
\begin{equation}
\begin{aligned}
\Delta_H&=1-\frac{24\lambda^4[\zeta'(\phi_H)]^2}{\epsilon r_H^4},\\
\phi'_H&=-\frac{6\lambda^2\zeta'(\phi_H)}{r_H^3(1+\sqrt{\Delta_H})},\qquad
b_1=\frac{1}{r_H+2\epsilon\lambda^2\zeta'(\phi_H)\phi'_H}.
\end{aligned}
\label{eq:poly-horizon}
\end{equation}
A real scalar slope requires $\Delta_H\geq0$, and $b_1>0$ ensures that
$B$ is positive just outside the horizon. In the canonical theory, the
condition $\Delta_H\geq0$ imposes a necessary bound on the domain of
existence of scalarized black holes \cite{Zou:2024paper}. For the phantom
theory, one has instead
\begin{equation}
\Delta_H=1+\frac{24\lambda^4[\zeta'(\phi_H)]^2}{r_H^4}\geq1,
\label{phantomdelta}
\end{equation}
with strict inequality for $\zeta'(\phi_H)\ne0$. Consequently, the
canonical horizon bound is absent in EPsGB theory. Exterior regularity
also restricts the solution families: some canonical branches terminate
at exterior curvature singularities with
$\Delta_H>0$ \cite{Zou:2024paper}.

At spatial infinity, asymptotic flatness requires
\begin{align}
A&=1-\frac{2M_A}{r}+O(r^{-2}),&
B&=1-\frac{2M_B}{r}+O(r^{-2}),&
\phi&=\frac{Q_s}{r}+O(r^{-2}).
\label{asymptotic}
\end{align}
Here $M_A$ and $M_B$ provide independent asymptotic estimates of the
ADM mass. We use $M\equiv M_A$, while $Q_s$ denotes the scalar charge.

With these boundary conditions, we solve Eq.~\eqref{eq:radial-Er}
algebraically for $A'/A$. Differentiating this expression and using
Eqs.~\eqref{eq:radial-Et} and \eqref{eq:radial-Ephi}, we obtain a linear
system for $B'$ and $\phi''$. For $\lambda=1$, this system is integrated
outward from the horizon expansions with an adaptive Runge--Kutta method,
together with $A$ and $\phi$.
Equation~\eqref{eq:radial-Eomega} is monitored as a constraint.

\begin{figure}[htbp]
\centering
\subfigure[$\zeta_p,\ \beta_p=125$]{\label{bg:z1phi}%
\includegraphics[width=0.325\textwidth]{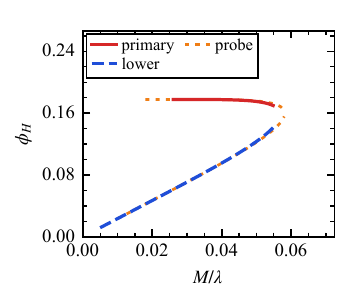}}%
\hfill%
\subfigure[$\zeta_p,\ \beta_p=0$]{\label{bg:z1phi0}%
\includegraphics[width=0.325\textwidth]{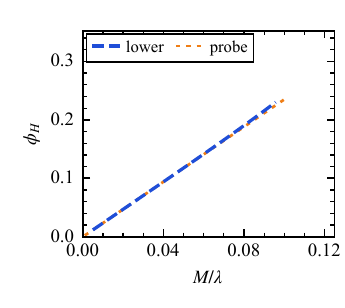}}%
\hfill%
\subfigure[$\zeta_e,\ \beta_e=1000$]{\label{bg:zephi}%
\includegraphics[width=0.325\textwidth]{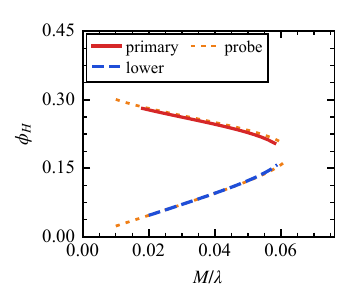}}%
\caption{Scalar field at the horizon versus mass for phantom black holes.
Solid and dashed curves denote the primary and lower branches, with
larger and smaller $\phi_H$ at fixed $r_H$ where both branches exist.
Dotted curves show Schwarzschild probe solutions.}
\label{bg:figstatic}
\end{figure}

The asymptotic scalar condition
$\phi_\infty=0$ is imposed by adjusting $r_H$ at fixed $\phi_H$, or
$\phi_H$ at fixed $r_H$. A rescaling of time then sets $A(\infty)=1$.
We retain nodeless, positive-mass solutions with regular horizons and
positive metric functions outside the horizon. The numerical accuracy
is tested for representative solutions by tightening the integration
tolerances and changing the asymptotic fits. Since both couplings are
even in $\phi$, we display only solutions with $\phi_H>0$.
In Fig.~\ref{bg:figstatic}, we show the scalar field at the horizon,
$\phi_H$, as a function of the mass $M$ for phantom black holes with
$\beta_p=125$, $\beta_p=0$ and $\beta_e=1000$.
At fixed $r_H$, we call the branches with larger and smaller $\phi_H$
the primary and lower branches, respectively.
The dotted curves show the corresponding Schwarzschild probe solutions;
the polynomial probe families were studied in Ref.~\cite{Zou:2024paper}.

The pure quartic solutions in Fig.~\ref{bg:z1phi0} lie on the lower
branch, along which $\phi_H$ increases almost linearly with $M$.
The lower branches for the other two
couplings approach the same curve for small $\phi_H$, where the quartic
term dominates.
By contrast, the primary branches in Figs.~\ref{bg:z1phi} and
\ref{bg:zephi} depend on the form of the coupling. For the polynomial
coupling, $\phi_H$ stays near the root $\phi_*$ of $\zeta_p'$ and changes
slowly with the mass. The exponential coupling has no nonzero root of
$\zeta_e'$, but the term
$\phi^3e^{-\beta_e\phi^4}$ in the source terms of the field equations is
exponentially suppressed for large $\phi$. Its primary branch lies above
the value $\phi_*$ of the polynomial coupling with $\beta_p=125$.
In both cases, backreaction shifts the branches away from the probe curves.

\begin{figure}[htbp]
\centering
\subfigure[$\zeta_p,\ \beta_p=125$]{\label{bg:metricz1}
\includegraphics[width=0.315\textwidth]{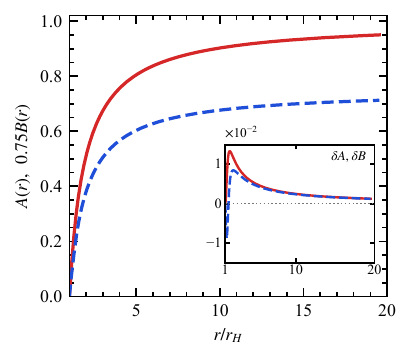}}
\hfill
\subfigure[$\zeta_p,\ \beta_p=0$]{\label{bg:metricz3}
\includegraphics[width=0.315\textwidth]{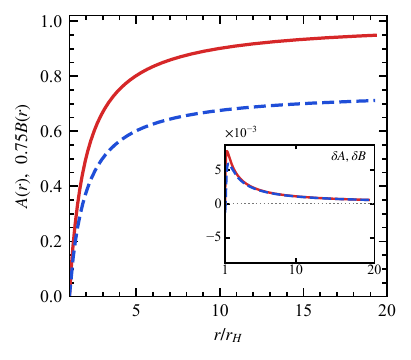}}
\hfill
\subfigure[$\zeta_e,\ \beta_e=1000$]{\label{bg:metricze}
\includegraphics[width=0.315\textwidth]{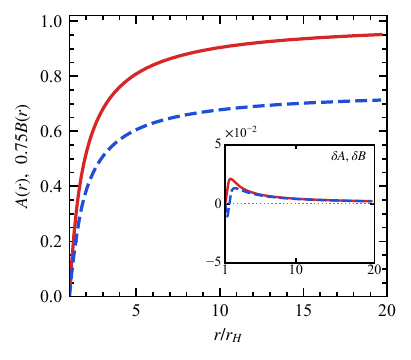}}
\caption{Phantom metric profiles at $r_H=0.10$.
Solid red and dashed blue curves show $A$ and $0.75B$, respectively;
the rescaling separates the curves visually.
Insets show the unscaled differences $\delta A=A-f_{\rm S}$ and
$\delta B=B-f_{\rm S}$, where $f_{\rm S}=1-r_H/r$ is the Schwarzschild
metric function for the same $r_H$.}
\label{bg:figgeometry}
\end{figure}

Figure~\ref{bg:figgeometry} shows representative metric profiles for
phantom black holes with $r_H=0.10$. The solutions in panels (a) and (c)
lie on the primary branches, and the pure quartic solution in panel
(b) lies on the lower branch. In all three cases, $A$ and $B$ vanish at
the horizon, remain positive outside it, and approach unity at infinity.
For visual separation, $B$ is plotted as $0.75B$. The insets show that
the differences from the Schwarzschild metric with the same $r_H$ tend
to zero at large radius. For these solutions, the Kretschmann scalar
remains finite outside the horizon and decays
asymptotically as $r^{-6}$.

\FloatBarrier

For the exponential coupling, our numerical survey covers canonical EsGB
and phantom EPsGB black holes with $\beta_e=25$, $100$ and $1000$.
In canonical EsGB theory, the horizon bound can separate distinct sets of
solutions \cite{Zou:2024paper,Blazquez-Salcedo:2022omw}. The canonical
solutions retained here satisfy $\Delta_H>0$. For $\beta_e=1000$, the
canonical and phantom solutions each form a continuous family over the
surveyed range. For $\beta_e=25$ and $100$, the phantom families also
remain continuous, and their mass tends toward zero at the largest
values of $\phi_H$ covered by the survey. Shooting for canonical
solutions fails to converge over some intervals of $\phi_H$.
The plots retain these gaps as numerically unresolved intervals.

\FloatBarrier

\section{Thermodynamics and phase structure}
\label{sec:thermodynamic-results}

The solutions constructed in Sec.~\ref{sec:numerical-solutions} are now
compared with Schwarzschild black holes through entropy at fixed mass
and free energy at fixed temperature \cite{Zou:2026thermo,Herdeiro:2026sur}.
Their heat capacities then test local stability against thermal fluctuations.

\subsection{Temperature, entropy and the first law}
\label{thermodynamics}
\label{sec:thermodynamic-quantities}
\label{sec:temperature-entropy}
\label{sec:entropy-ordering}
\label{sec:poly-diagrams}
\label{sec:exp-thermodynamics}
\label{sec:coupling-comparison}
\label{sec:entropy-comparison}
\label{sec:exp-first-law}

With the time coordinate normalized by $A(\infty)=1$, the near-horizon
expansions in Eq.~\eqref{horizon} give the Hawking temperature in terms of
the surface gravity $\kappa$,
\begin{equation}
T_H=\frac{\kappa}{2\pi}
 =\frac{1}{4\pi}\lim_{r\to r_H}\sqrt{\frac{B(r)}{A(r)}}\,A'(r)
 =\frac{\sqrt{a_1b_1}}{4\pi}.
\label{temperature}
\end{equation}
The Wald entropy includes the area and Gauss--Bonnet contributions
\cite{Wald:1993nt,Iyer:1994ys},
\begin{equation}
S_H=\pi r_H^2+4\pi\epsilon\lambda^2\zeta(\phi_H).
\label{thermo}
\end{equation}
The correction carries the sign $\epsilon$ of the Gauss--Bonnet term
in the action. For phantom black holes, $\epsilon=-1$ gives
\begin{equation}
S_H=\pi r_H^2-4\pi\lambda^2\zeta(\phi_H).
\label{phantomentropy}
\end{equation}
The reduced temperature and entropy compare each solution with
Schwarzschild at the same mass,
\begin{equation}
t_H=8\pi MT_H,\qquad
s_H=\frac{S_H}{4\pi M^2}.
\label{eq:exp-reduced}
\end{equation}
so that $t_H=s_H=1$ for Schwarzschild. All quantities below are
expressed in units of $\lambda$. For the polynomial coupling with
$\beta_p=125$ and the exponential coupling, the horizon-scalar coordinate is
\begin{equation}
x=\begin{cases}
\phi_H/\phi_*, & \text{polynomial coupling with }\beta_p=125,\\
\beta_e^{1/4}\phi_H, & \text{exponential coupling},
\end{cases}
\label{eq:poly-parameter}
\end{equation}
where $\phi_*$ is defined in Eq.~\eqref{roots}. These coordinates coincide
for $(\beta_p,\beta_e)=(125,1000)$. The pure quartic family, $\beta_p=0$,
is parametrized directly by $\phi_H$.

\begin{figure}[htbp]
\centering
\subfigure[$t_H$, $\beta_p=125$]{\label{fig:poly-reduced-t125}%
\includegraphics[width=0.485\textwidth]{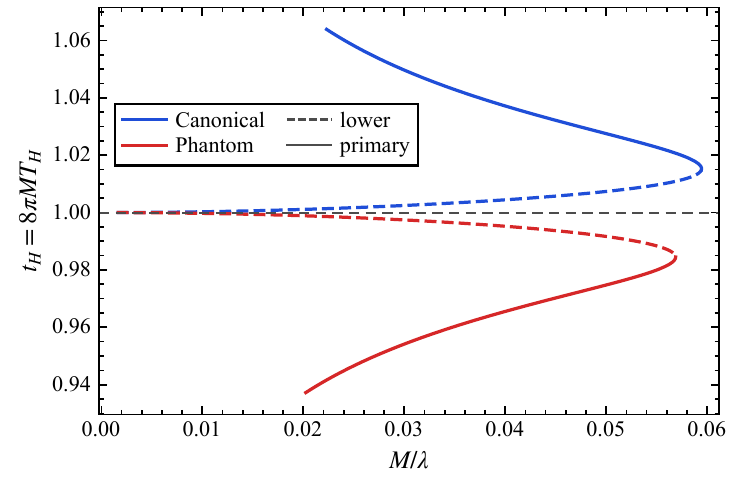}}%
\hfill%
\subfigure[$s_H$, $\beta_p=125$]{\label{fig:poly-reduced-s125}%
\includegraphics[width=0.485\textwidth]{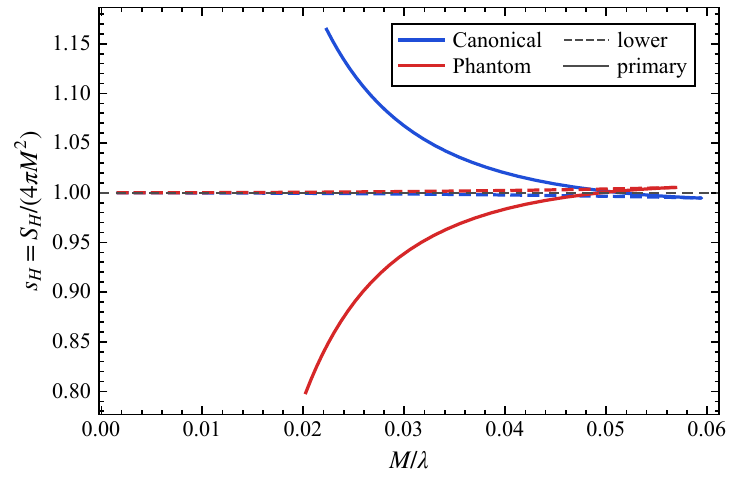}}%
\par
\subfigure[$t_H$, $\beta_p=0$]{\label{fig:poly-reduced-t0}%
\includegraphics[width=0.485\textwidth]{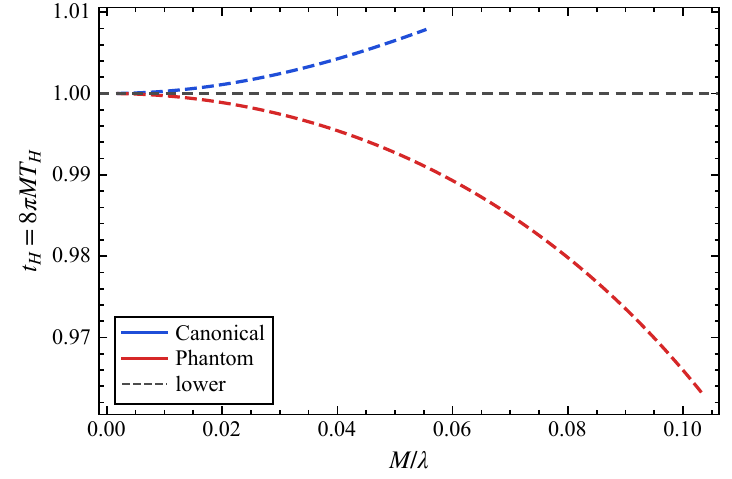}}%
\hfill%
\subfigure[$s_H$, $\beta_p=0$]{\label{fig:poly-reduced-s0}%
\includegraphics[width=0.485\textwidth]{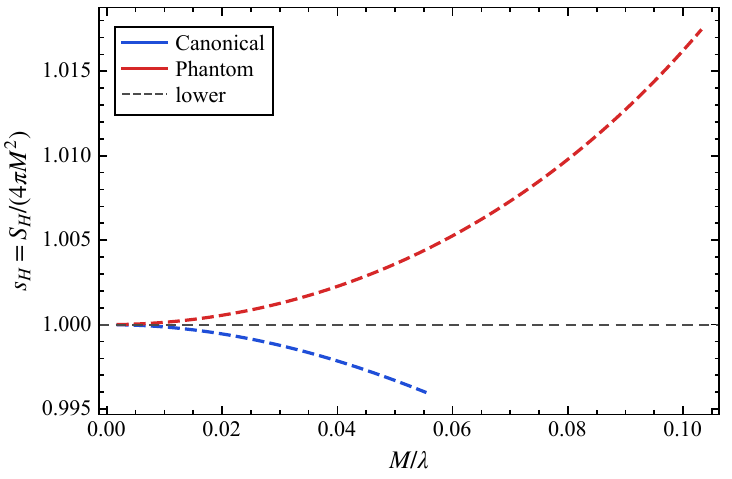}}%
\caption{Reduced temperature and entropy versus mass for the polynomial coupling,
with $\beta_p=125$ (top) and $0$ (bottom). Blue: canonical; red: phantom.
Dashed and solid curves denote the lower and primary branches.
The horizontal reference lines give the Schwarzschild values $t_H=s_H=1$.}
\label{fig:poly-reduced}
\end{figure}

Figure~\ref{fig:poly-reduced} compares the reduced temperature and
entropy for polynomial black holes. On the lower branches, phantom
solutions have $s_H>1$ and canonical solutions have $s_H<1$.
For $\beta_p=125$, the phantom primary branch crosses from $s_H>1$ to
$s_H<1$ as $\phi_H$ increases; the canonical primary branch has the
opposite ordering.
The pure quartic curves stay on opposite sides of $s_H=1$ over the range shown.
In this pure quartic limit, canonical and phantom temperatures satisfy
$t_H>1$ and $t_H<1$, respectively [Fig.~\ref{fig:poly-reduced-t0}].

\FloatBarrier

\begin{figure}[htbp]
\centering
\subfigure[$t_H$, canonical]{\label{fig:exp-reduced-t}%
\includegraphics[width=0.485\textwidth]{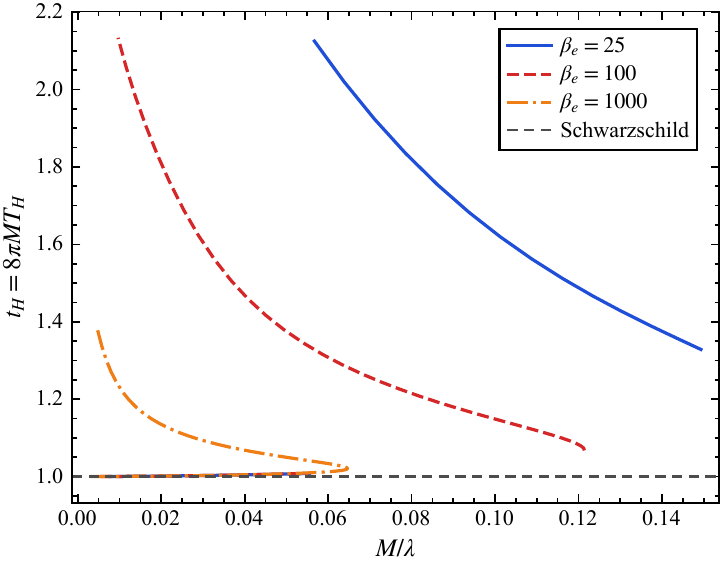}}%
\hfill%
\subfigure[$s_H$, canonical]{\label{fig:exp-reduced-s}%
\includegraphics[width=0.485\textwidth]{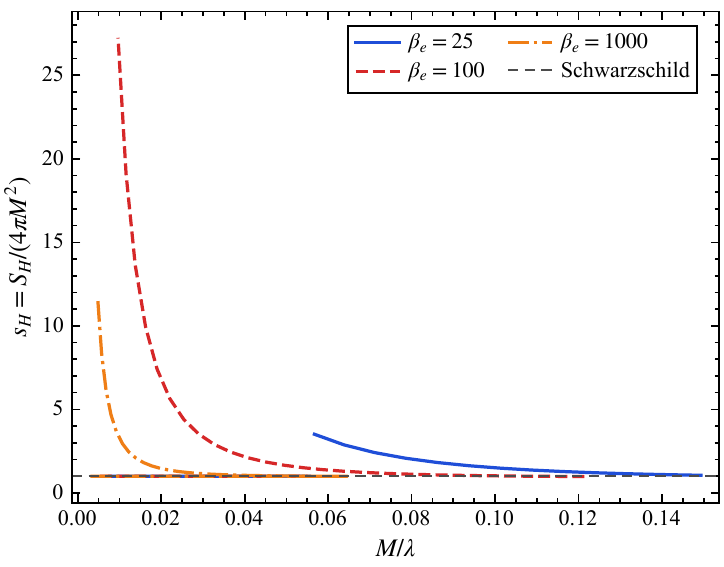}}%
\par
\subfigure[$t_H$, phantom]{\label{fig:exp-reduced-t-phantom}%
\includegraphics[width=0.485\textwidth]{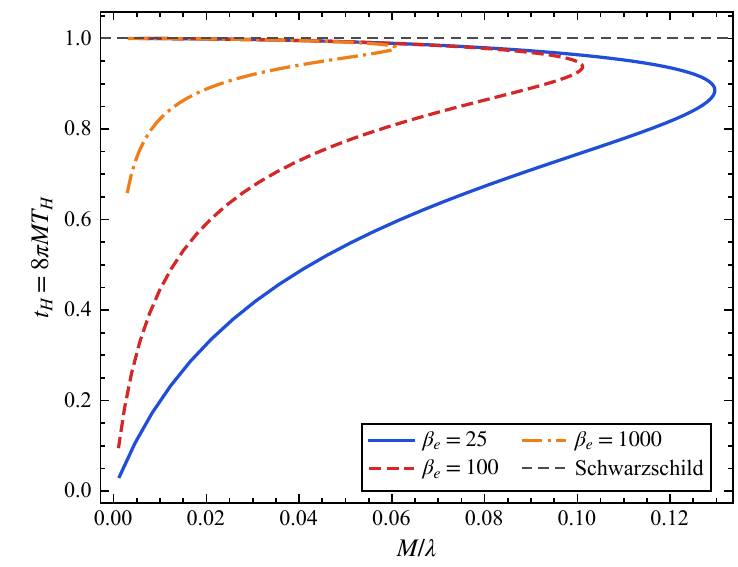}}%
\hfill%
\subfigure[$s_H$, phantom]{\label{fig:exp-reduced-s-phantom}%
\includegraphics[width=0.485\textwidth]{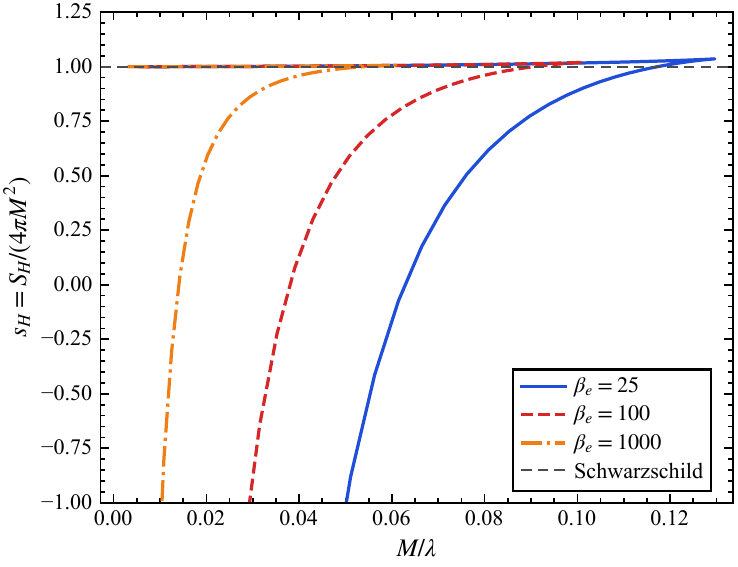}}%
\caption{Reduced temperature and entropy versus mass for canonical (top) and
phantom (bottom) exponential black holes. Line styles distinguish the
values of $\beta_e$, and horizontal reference lines give $t_H=s_H=1$.
Gaps denote unresolved numerical intervals. Panel (d) is truncated at
$s_H=-1$ to resolve the entropy zero crossings.}
\label{fig:exp-reduced}
\end{figure}

The exponential families have the same lower-branch entropy ordering
[Fig.~\ref{fig:exp-reduced}]. Their temperatures satisfy $t_H>1$ in the
canonical theory and $t_H<1$ in the phantom theory.
The entropy ordering persists on parts of the primary branches and
reverses at larger $\phi_H$.
On the phantom primary branches, $s_H$ decreases with decreasing mass
and becomes negative. Equation~\eqref{phantomentropy} gives the condition
$r_H^2<4\lambda^2\zeta_e(\phi_H)$.
Since $\zeta_e\leq1/(4\beta_e)$, negative entropy requires
$r_H<\lambda\beta_e^{-1/2}$.

\FloatBarrier

The temperature and entropy are checked against the first law at fixed
$\lambda$, $\epsilon$, $\beta_p$ or $\beta_e$, and $\phi_\infty=0$.
The first law and its relative numerical residual are
\begin{align}
\ud M&=T_H\ud S_H,
\label{firstlaw}\\
\mathcal E_{\rm FL}&=\left|1-T_H\frac{\ud S_H}{\ud M}\right|.
\label{firstlawresidual}
\end{align}
Derivatives are evaluated by finite differences within continuous
segments of the nonuniform solution grid, away from endpoints and mass
extrema.
For representative polynomial solutions, $\mathcal E_{\rm FL}$ ranges
from $3.83\times10^{-8}$ to $5.24\times10^{-7}$.
For the exponential coupling, the ranges are $3.50\times10^{-8}$ to
$3.87\times10^{-7}$ for canonical solutions and $4.02\times10^{-8}$ to
$2.44\times10^{-7}$ for phantom solutions.

\begin{figure}[htbp]
\centering
\subfigure[$\zeta_p,\ \beta_p=125$]{\includegraphics[width=0.485\textwidth]{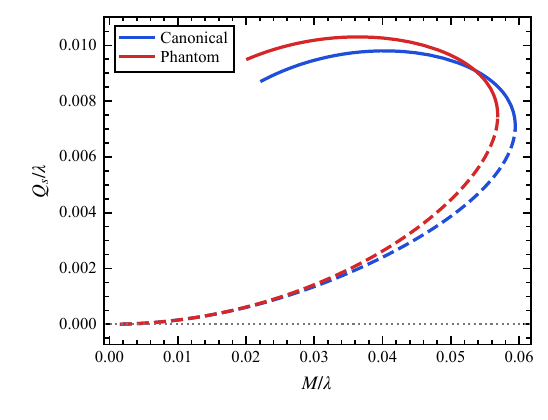}}
\hfill
\subfigure[$\zeta_p,\ \beta_p=0$]{\includegraphics[width=0.485\textwidth]{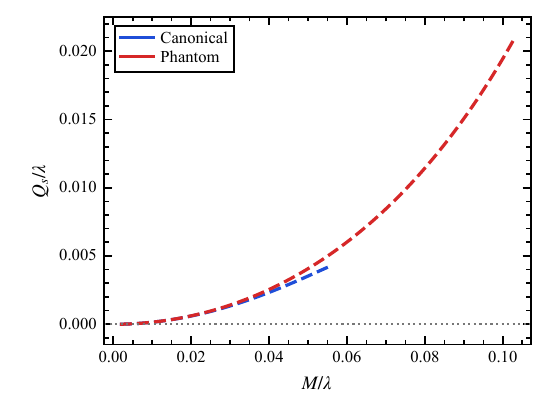}}
\caption{Scalar charge versus mass for the polynomial coupling with
$\beta_p=125$ (left) and $0$ (right). Blue: canonical; red: phantom.
Dashed and solid curves denote the lower and primary branches.}
\label{fig:poly-charge}
\end{figure}

The scalar charge $Q_s$ is secondary hair and does not enter
Eq.~\eqref{firstlaw} as an independent thermodynamic variable. Its
variation along the canonical and phantom branches is shown in
Fig.~\ref{fig:poly-charge}; the canonical branches were studied in
Ref.~\cite{Zou:2024paper}. For $\beta_p=125$, the lower and primary
branches join at the maximum of $r_H$ and turn back in mass, allowing
several solutions with the same $M$.
For $\beta_p=0$, the phantom branch extends to larger masses than the
canonical branch in the computed range. The latter terminates at the
horizon bound $\Delta_H=0$, which is absent in the phantom theory
[Eq.~\eqref{phantomdelta}].

\FloatBarrier

\begin{figure}[htbp]
\centering
\subfigure[$M$]{\label{fig:matched-couplings-mass}%
\includegraphics[width=0.485\textwidth]{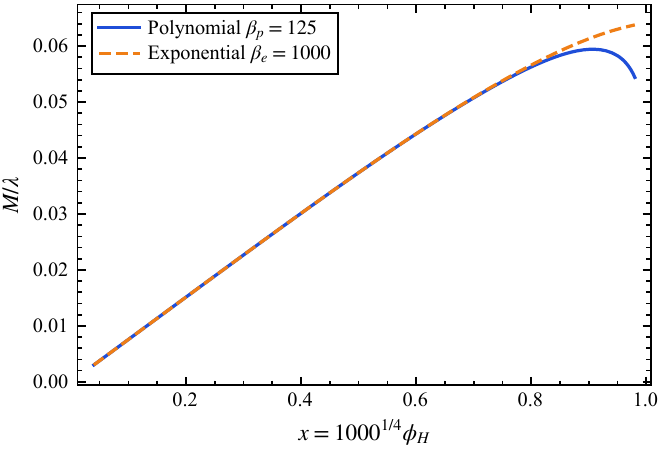}}%
\hfill%
\subfigure[$T_H$]{\label{fig:matched-couplings-temperature}%
\includegraphics[width=0.485\textwidth]{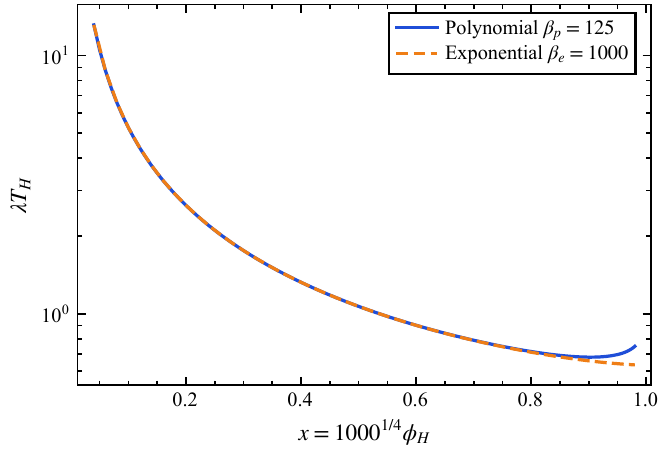}}%
\caption{Canonical mass (a) and temperature (b) for the polynomial coupling
$\beta_p=125$ and exponential coupling $\beta_e=1000$, compared at the
same $x=1000^{1/4}\phi_H$. The temperature axis is logarithmic.
The interval $0.04\le x\le0.98$ includes the polynomial extrema;
the exponential extrema lie beyond it.}
\label{fig:matched-couplings}
\end{figure}

For comparison, we also consider the canonical solutions for the polynomial
and exponential couplings. Their small-field expansions agree
through order $\phi^8$ when
\begin{equation}
\beta_p=\frac{\beta_e}{8},\qquad
\zeta_e-\zeta_p=\frac{\beta_e^2\phi^{12}}{24}+O(\phi^{16}).
\label{eq:coupling-match}
\end{equation}
For $(\beta_p,\beta_e)=(125,1000)$, both families use
$x=1000^{1/4}\phi_H$. Their nonlinear source terms differ at finite field,
\begin{equation}
u=\beta_e\phi_H^4=x^4,\qquad
\zeta'_p(\phi)=\phi^3(1-\beta_e\phi^4),\qquad
\zeta'_e(\phi)=\phi^3e^{-\beta_e\phi^4}.
\label{eq:expansion-domain}
\end{equation}
The polynomial derivative changes sign at $\phi_*=1000^{-1/4}$.
The exponential derivative remains positive for finite $\phi>0$ and is
exponentially suppressed at large field.
Figure~\ref{fig:matched-couplings} shows that mass and temperature nearly
coincide at small $x$ but separate as $\phi_H$ increases.
The polynomial extrema lie within the plotted interval; the exponential
extrema lie beyond it.

\FloatBarrier

\begin{figure}[htbp]
\centering
\subfigure[$\zeta_p,\ \beta_p=125$]{\label{fig:matched-entropy-poly}%
\includegraphics[width=0.485\textwidth]{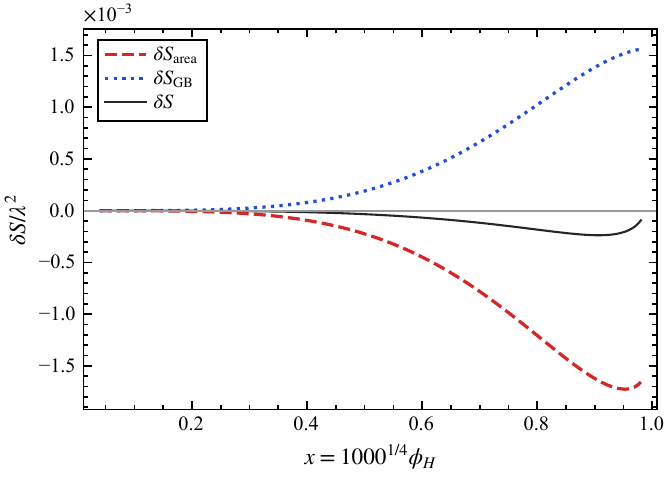}}%
\hfill%
\subfigure[$\zeta_e,\ \beta_e=1000$]{\label{fig:matched-entropy-exp}%
\includegraphics[width=0.485\textwidth]{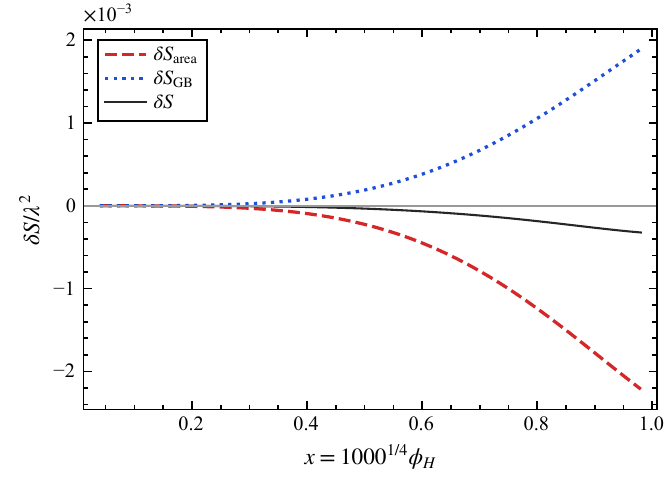}}%
\caption{Entropy difference from Schwarzschild at the same mass for canonical
polynomial (a) and exponential (b) black holes, with
$(\beta_p,\beta_e)=(125,1000)$. The red dashed and blue dotted curves
give the area and Gauss--Bonnet contributions; the solid black curve
gives their sum. Both families use $x=1000^{1/4}\phi_H$ over
$0.04\le x\le0.98$.}
\label{fig:matched-entropy}
\end{figure}

The entropy difference from Schwarzschild at the same mass,
$\delta S=S_H-4\pi M^2$, separates into area and Gauss--Bonnet terms,
\begin{equation}
\delta S=\underbrace{\pi(r_H^2-4M^2)}_{\delta S_{\rm area}}
       +\underbrace{4\pi\epsilon\lambda^2\zeta(\phi_H)}_{\delta S_{\rm GB}}.
\label{eq:entropy-decomposition}
\end{equation}
Figure~\ref{fig:matched-entropy} shows their competition for the two
canonical families. At $x=0.98$, the polynomial contributions in units
of $\lambda^2$ are $-1.65525\times10^{-3}$ and $1.56133\times10^{-3}$,
respectively; the exponential values are $-2.21466\times10^{-3}$ and
$1.89257\times10^{-3}$. The area deficit outweighs the positive
Gauss--Bonnet correction in both cases, giving $\delta S<0$.
Changing $\epsilon$ also changes the metric backreaction, so the
phantom entropy reversal involves both terms in
Eq.~\eqref{eq:entropy-decomposition}.

\FloatBarrier

\subsection{Heat capacity and local stability}
\label{sec:thermal-response}
\label{sec:poly-response}
\label{sec:exp-capacity}

We examine local thermal stability in the canonical ensemble for both
canonical EsGB ($\epsilon=+1$) and phantom EPsGB ($\epsilon=-1$)
black holes. The ensemble fixes the temperature; the labels canonical
and phantom refer to the sign of the scalar kinetic term.
The heat capacity at fixed coupling parameters is
\begin{equation}
C=\frac{\ud M}{\ud T_H}
 =\frac{\ud M/\ud x}{\ud T_H/\ud x},
\label{heatcapacity}
\end{equation}
with $\phi_H$ replacing $x$ for $\beta_p=0$.
Let $x_T$ and $x_M$ denote the temperature minimum and mass maximum.
At $x_T$, $C$ diverges because $\ud T_H/\ud x=0$ and $\ud M/\ud x\ne0$;
at $x_M$, it vanishes because $\ud M/\ud x=0$ and $\ud T_H/\ud x\ne0$.
The first law also implies $\ud S_H/\ud x=0$ at $x_M$.
Near these extrema, the sign of $C$ is determined from the two derivatives
in Eq.~\eqref{heatcapacity}.
The refined integrations use tighter tolerances, a starting point closer
to the horizon, a larger outer radius and a higher-order asymptotic fit.
The sign is unchanged between the two settings and under variations
of the derivative spacing.

\begin{figure}[htbp]
\centering
\subfigure[$M$, canonical]{\label{fig:poly-turn1-mass-canonical}%
\includegraphics[width=0.325\textwidth]{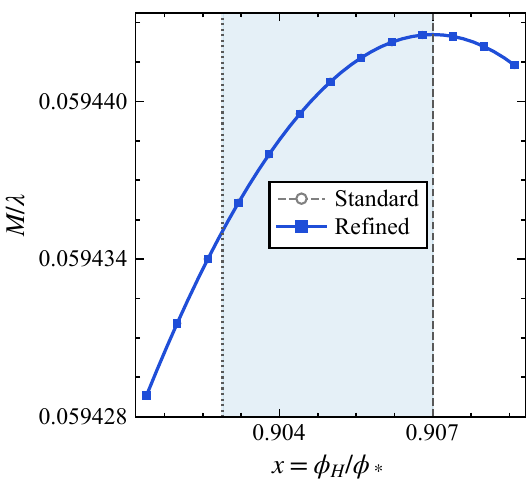}}%
\hfill%
\subfigure[$T_H$, canonical]{\label{fig:poly-turn1-temperature-canonical}%
\includegraphics[width=0.325\textwidth]{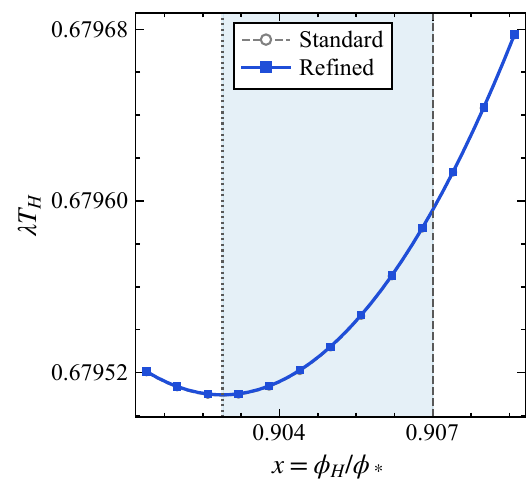}}%
\hfill%
\subfigure[$C$, canonical]{\label{fig:poly-turn1-capacity-canonical}%
\includegraphics[width=0.325\textwidth]{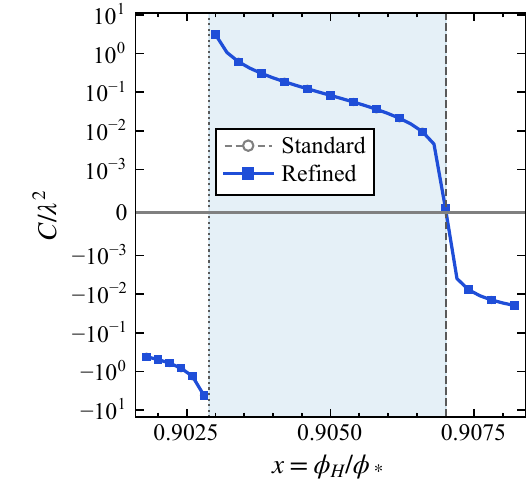}}%
\par
\subfigure[$M$, phantom]{\label{fig:poly-turn1-mass-phantom}%
\includegraphics[width=0.325\textwidth]{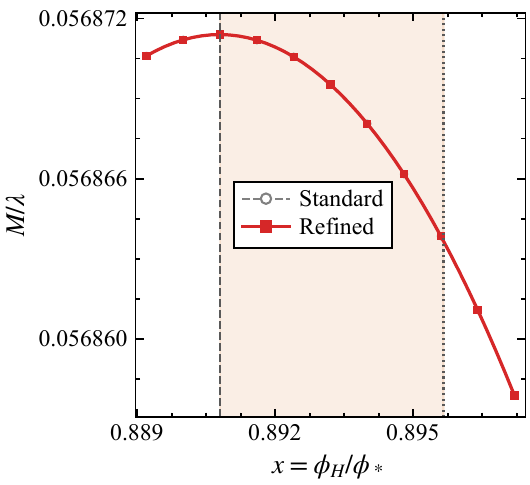}}%
\hfill%
\subfigure[$T_H$, phantom]{\label{fig:poly-turn1-temperature-phantom}%
\includegraphics[width=0.325\textwidth]{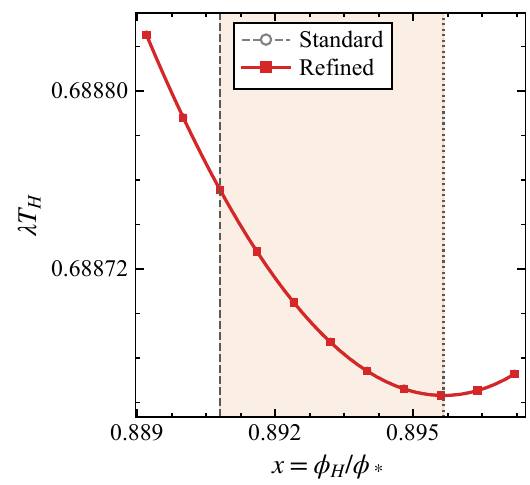}}%
\hfill%
\subfigure[$C$, phantom]{\label{fig:poly-turn1-capacity-phantom}%
\includegraphics[width=0.325\textwidth]{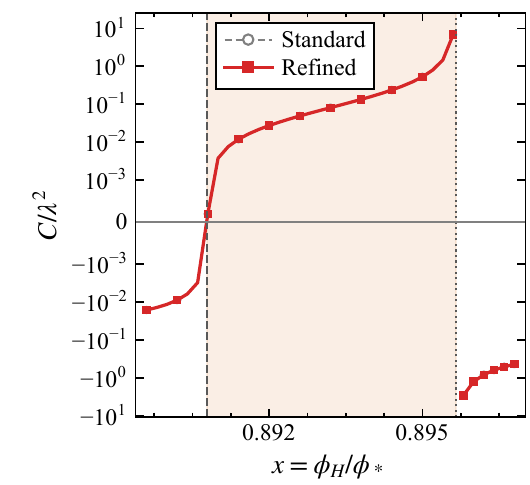}}%
\caption{Mass, temperature and heat capacity near the extrema for canonical
(top) and phantom (bottom) polynomial black holes with $\beta_p=125$.
Circles and squares compare the standard and refined integration settings.
Dotted and dashed lines mark $x_T$ and $x_M$; shading marks $C>0$.
Heat capacities use a symmetric logarithmic scale, with separate curves
on the two sides of each divergence.}
\label{fig:poly-turn1}
\end{figure}

Figure~\ref{fig:poly-turn1} resolves the polynomial extrema for $\beta_p=125$.
Canonical solutions have $x_T<x_M$, with both derivatives positive
between the extrema. Phantom solutions have $x_M<x_T$, with both negative.
Canonical EsGB and phantom EPsGB black holes therefore have $C>0$
between the extrema, with opposite orderings. The interval widths
in $x$ are $0.00412$ and $0.00487$ for
canonical and phantom black holes, respectively.

\FloatBarrier

\begin{figure}[htbp]
\centering
\subfigure[$\beta_p=125$]{\includegraphics[width=0.485\textwidth]{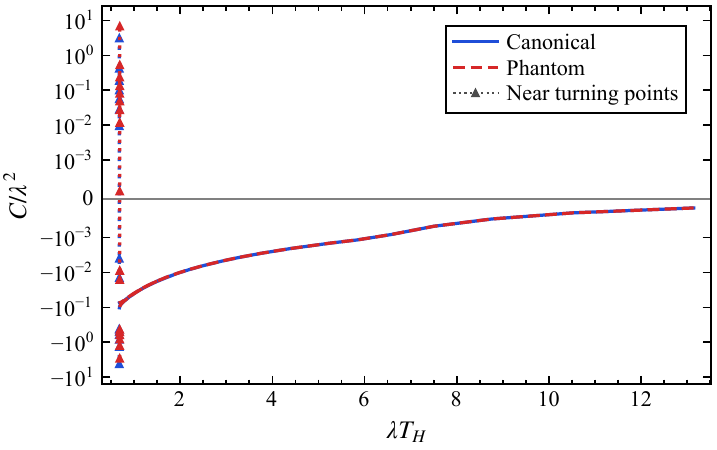}}
\hfill
\subfigure[$\beta_p=0$]{\includegraphics[width=0.485\textwidth]{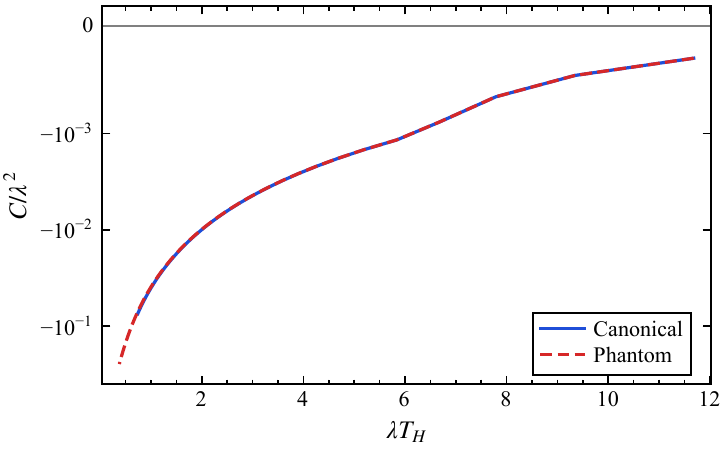}}
\caption{Heat capacity versus temperature for polynomial black holes.
Blue: canonical; red: phantom. Triangles resolve the turning regions for
$\beta_p=125$. The vertical axes use a symmetric logarithmic scale,
which retains both signs and is linear near $C=0$. Curves are separated
at the temperature minima, where $C$ diverges.}
\label{fig:poly-capacity}
\end{figure}

Figure~\ref{fig:poly-capacity} shows the polynomial heat capacities versus
temperature. The refined turning-point data resolve the narrow positive
intervals for $\beta_p=125$. The pure quartic curves have $C<0$
throughout the range shown.

\FloatBarrier

\begin{figure}[htbp]
\centering
\subfigure[$M$]{\label{fig:exp-turn-mass}%
\includegraphics[width=0.325\textwidth]{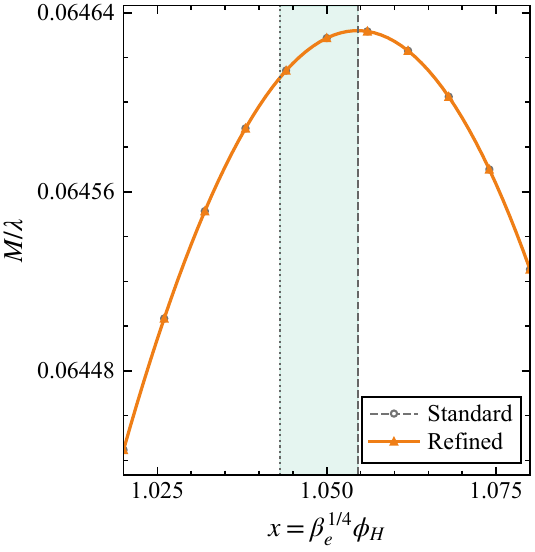}}%
\hfill%
\subfigure[$T_H$]{\label{fig:exp-turn-temperature}%
\includegraphics[width=0.325\textwidth]{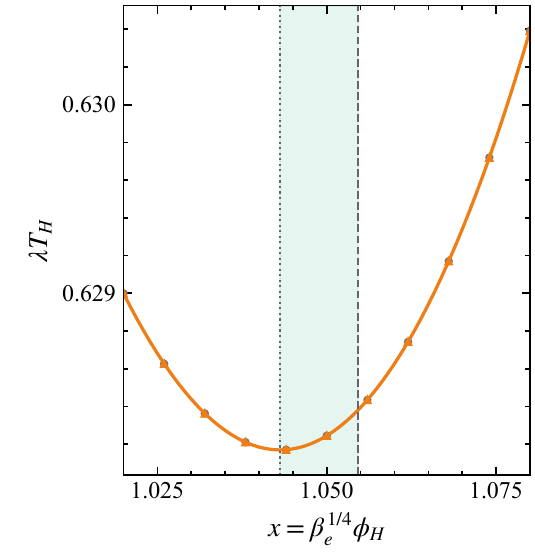}}%
\hfill%
\subfigure[$C$]{\label{fig:exp-turn-capacity}%
\includegraphics[width=0.325\textwidth]{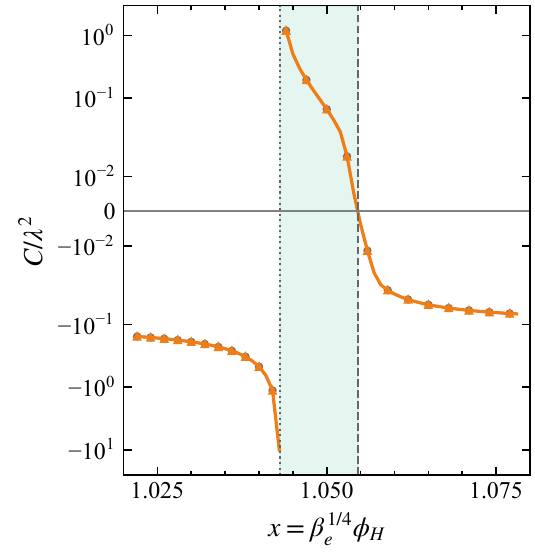}}%
\caption{Mass, temperature and heat capacity near the extrema for canonical
exponential black holes with $\beta_e=1000$. The curves compare the
standard and refined integration settings. Dotted and dashed lines mark
$x_T$ and $x_M$; shading marks $C>0$. Heat capacity uses a symmetric
logarithmic scale and is separated at its divergence.}
\label{fig:exp-turn}
\end{figure}

The canonical exponential family with $\beta_e=1000$ also has $x_T<x_M$.
Figure~\ref{fig:exp-turn} resolves the extrema at $x_T\simeq1.0431$ and
$x_M\simeq1.0546$. Both mass and temperature increase between them,
giving $C>0$.

\FloatBarrier

\begin{figure}[!htbp]
\centering
\subfigure[$M$, $\beta_e=25$]{\label{fig:exp-turn-phantom-mass25}%
\includegraphics[width=0.325\textwidth]{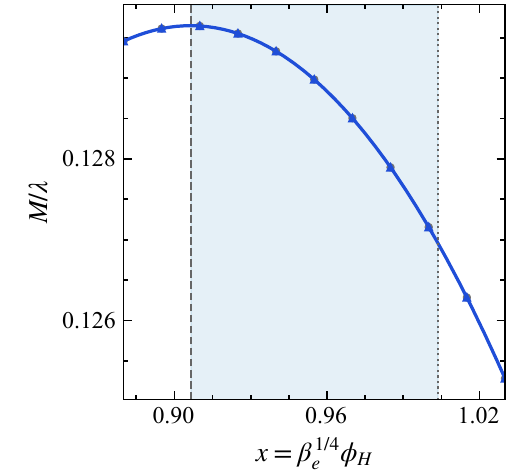}}%
\hfill%
\subfigure[$T_H$, $\beta_e=25$]{\label{fig:exp-turn-phantom-temperature25}%
\includegraphics[width=0.325\textwidth]{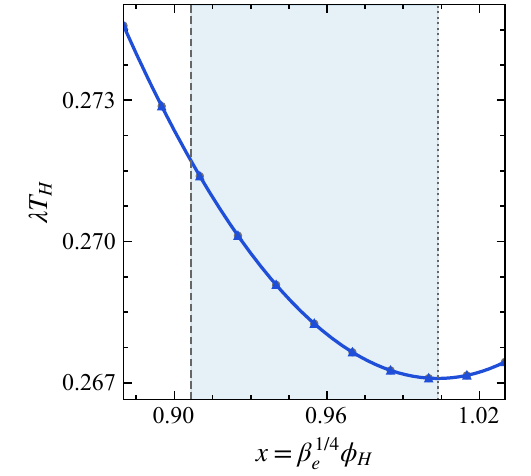}}%
\hfill%
\subfigure[$C$, $\beta_e=25$]{\label{fig:exp-turn-phantom-capacity25}%
\includegraphics[width=0.325\textwidth]{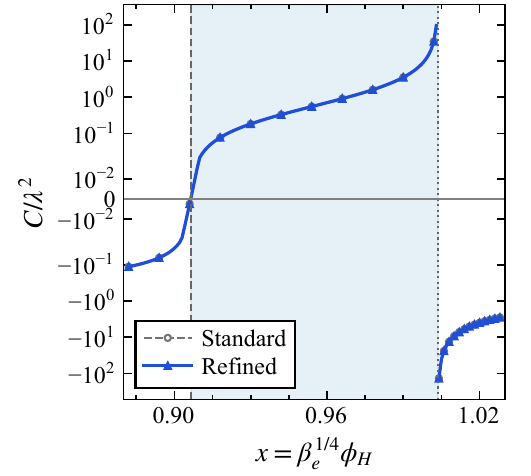}}%
\par
\subfigure[$M$, $\beta_e=100$]{\label{fig:exp-turn-phantom-mass100}%
\includegraphics[width=0.325\textwidth]{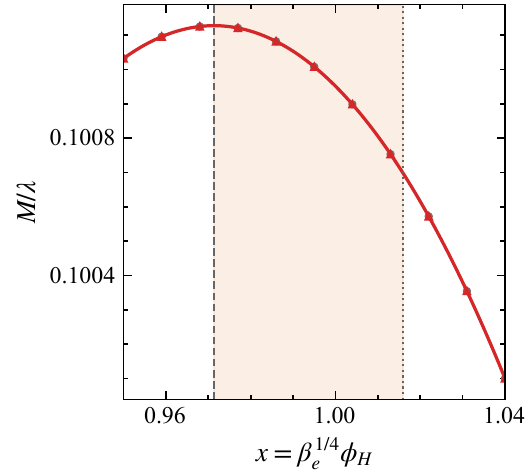}}%
\hfill%
\subfigure[$T_H$, $\beta_e=100$]{\label{fig:exp-turn-phantom-temperature100}%
\includegraphics[width=0.325\textwidth]{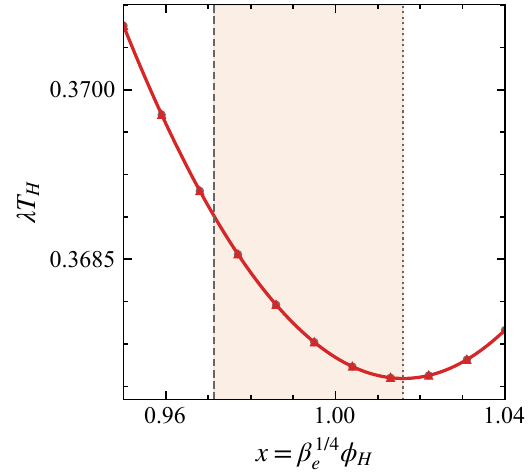}}%
\hfill%
\subfigure[$C$, $\beta_e=100$]{\label{fig:exp-turn-phantom-capacity100}%
\includegraphics[width=0.325\textwidth]{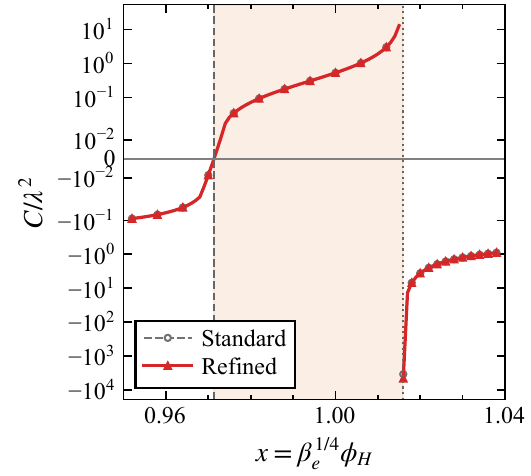}}%
\par
\subfigure[$M$, $\beta_e=1000$]{\label{fig:exp-turn-phantom-mass1000}%
\includegraphics[width=0.325\textwidth]{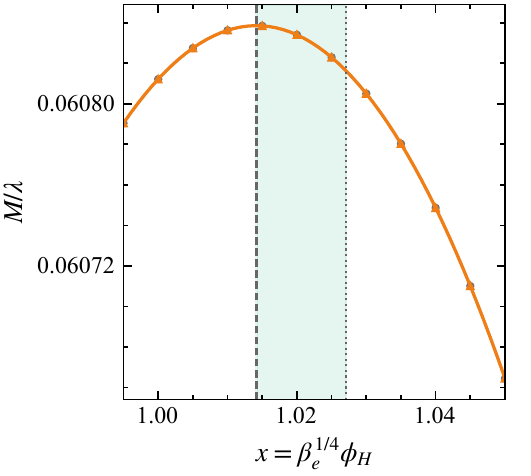}}%
\hfill%
\subfigure[$T_H$, $\beta_e=1000$]{\label{fig:exp-turn-phantom-temperature1000}%
\includegraphics[width=0.325\textwidth]{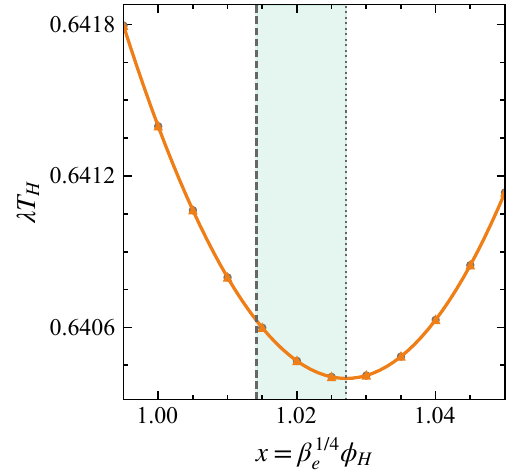}}%
\hfill%
\subfigure[$C$, $\beta_e=1000$]{\label{fig:exp-turn-phantom-capacity1000}%
\includegraphics[width=0.325\textwidth]{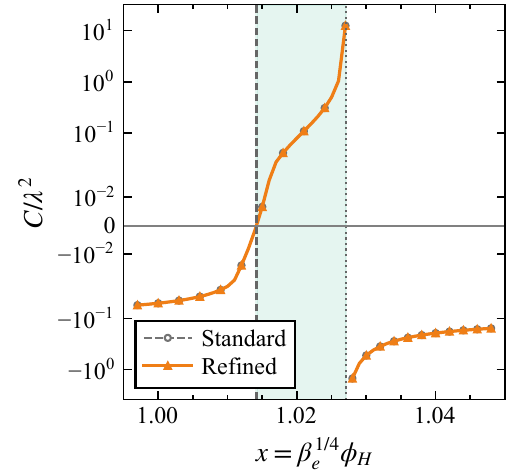}}%
\caption{Mass, temperature and heat capacity near the extrema for phantom
exponential black holes, with $\beta_e=25$, $100$ and $1000$ from top
to bottom. Circles and triangles compare the standard and refined
integration settings. Dashed and dotted lines mark $x_M$ and $x_T$;
shading marks $C>0$. Heat capacities use symmetric logarithmic scales
and are separated at their divergences.}
\label{fig:exp-turn-phantom}
\end{figure}

For the phantom exponential families, the order is reversed to $x_M<x_T$
[Fig.~\ref{fig:exp-turn-phantom}]. Mass and temperature both decrease
between the extrema, again giving $C>0$.
The interval widths are $0.0972$, $0.0447$ and $0.0129$ for
$\beta_e=25$, $100$ and $1000$, respectively.
The remaining segments of the canonical and phantom exponential families
with resolved derivatives have negative heat capacity.
Table~\ref{tab:poly-turns} collects the extrema bounding the positive intervals.

\begin{table}[htbp]
\centering\small
\caption{Temperature minima and mass maxima bounding the regions of positive heat capacity.}
\label{tab:poly-turns}
\begin{tabular}{llrrr}
\toprule
Coupling & Parameter & $\epsilon$ & $x_T$ & $x_M$ \\
\midrule
$\zeta_p$ & $\beta_p=125$ & 1 & 0.90289 & 0.90700 \\
$\zeta_p$ & $\beta_p=125$ & -1 & 0.89566 & 0.89079 \\
$\zeta_e$ & $\beta_e=1000$ & 1 & 1.0431 & 1.0546 \\
$\zeta_e$ & $\beta_e=25$ & -1 & 1.0036 & 0.9064 \\
$\zeta_e$ & $\beta_e=100$ & -1 & 1.0160 & 0.9713 \\
$\zeta_e$ & $\beta_e=1000$ & -1 & 1.0271 & 1.0142 \\
\bottomrule
\end{tabular}
\end{table}

\FloatBarrier

The heat capacity characterizes the local thermal response.
Perturbation spectra and hyperbolicity describe the dynamical properties
\cite{Blazquez-Salcedo:2018jnn,Silva:2018qhn,
Blazquez-Salcedo:2020rhf,Blazquez-Salcedo:2020caw,Herdeiro:2026sur}.

\FloatBarrier

\subsection{Free energy and thermodynamic crossings}
\label{sec:free-energy-results}
\label{sec:exp-ensemble}
\label{sec:exp-order-parameter}
\label{sec:free-energy-crossings}
\label{sec:poly-crossings}
\label{sec:exp-crossings}
\label{sec:interpretation}

At fixed temperature, the Helmholtz free energy is $F=M-T_HS_H$.
The Schwarzschild reference quantities are
\begin{equation}
T_{\rm S}=\frac{1}{8\pi M_{\rm S}},\qquad
S_{\rm S}=4\pi M_{\rm S}^2,\qquad
F_{\rm S}=\frac{1}{16\pi T_{\rm S}}.
\label{schwthermo}
\end{equation}
The differences relevant to entropy and free-energy ordering are
\begin{equation}
\delta S=S_H-4\pi M^2,\qquad
\delta F=F-\frac{1}{16\pi T_H}.
\label{thermodifferences}
\end{equation}
Here $\delta S$ compares solutions at equal mass and $\delta F$
compares them at equal temperature. Using the coordinate $x$ in
Eq.~\eqref{eq:poly-parameter}, the crossings satisfy \cite{Zou:2026thermo}
\begin{equation}
\delta S(x_S)=0,\qquad \delta F(x_F)=0.
\label{eq:exp-crossing-conditions}
\end{equation}
Each root is found by bisection in $x$, solving the field equations at
every step. The two conditions generally select different states, with
common mass $M_S$ or temperature $T_F$.
At $T_F$, the scalarized mass $M_F$ and the entropy give the differences
\begin{equation}
\begin{aligned}
\Delta S_{T_F}&=S_H(T_F)-\frac{1}{16\pi T_F^2},\\
\Delta M_{T_F}&=M_F-\frac{1}{8\pi T_F}=T_F\Delta S_{T_F}.
\end{aligned}
\label{eq:poly-jumps}
\end{equation}
The final equality follows from $\delta F(T_F)=0$.

\begin{figure}[htbp]
\centering
\subfigure[$\beta_p=125$]{\label{fig:poly-free-delta125}%
\includegraphics[width=0.485\textwidth]{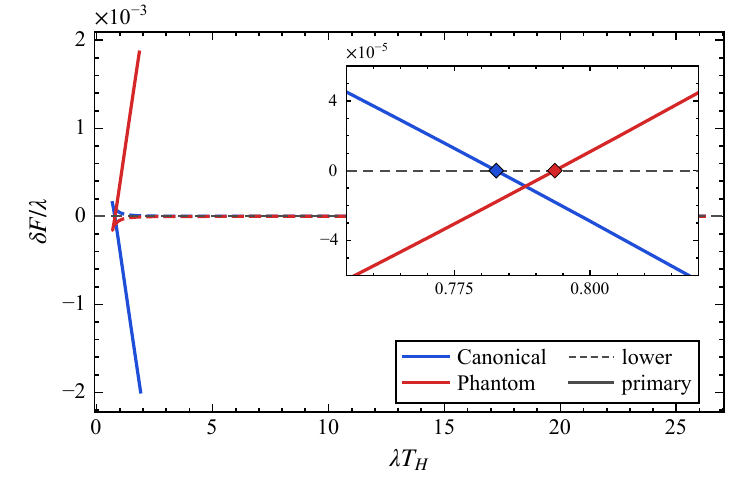}}%
\hfill
\subfigure[$\beta_p=0$]{\label{fig:poly-free-delta0}%
\includegraphics[width=0.485\textwidth]{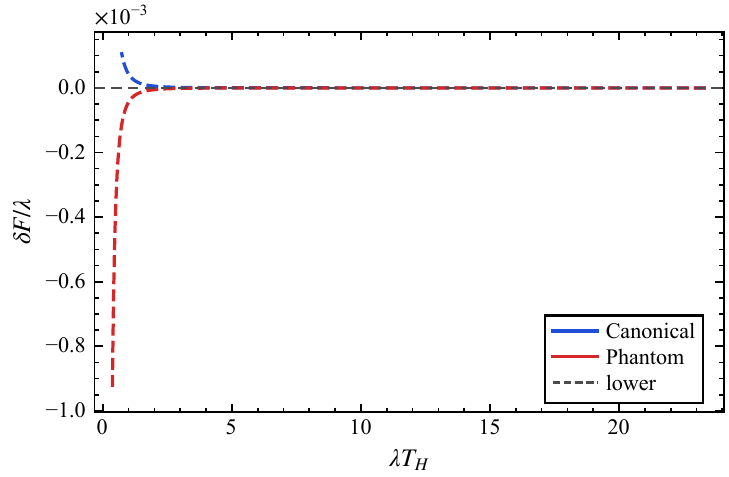}}%
\caption{Free-energy difference from Schwarzschild at the same
temperature for polynomial black holes. Blue: canonical; red: phantom.
Dashed and solid curves denote the lower and primary branches.
The inset in (a) resolves the crossings, marked by diamonds.
The horizontal reference line is $\delta F=0$.}
\label{fig:poly-free}
\end{figure}

Figure~\ref{fig:poly-free} compares the polynomial free-energy differences.
On the lower branches, $\delta F>0$ for canonical solutions
and $\delta F<0$ for phantom solutions.
For $\beta_p=125$, the phantom primary branch has $\delta F<0$ for
$x<x_F$ and $\delta F>0$ for $x>x_F$; the canonical primary branch has
the opposite signs.
The roots are $x_F\simeq0.987545$ and $0.988320$ for canonical and
phantom black holes; the entropy roots are $x_S\simeq0.989107$ and
$0.986542$, respectively. Table~\ref{tab:poly-crossings} gives the
corresponding masses, temperatures and jumps.
In the pure quartic limit, the lower-branch ordering persists throughout
the computed range, without a free-energy crossing.

\begin{table}[htbp]
\centering\small
\caption{Entropy crossing points at mass $M_S$ and free-energy crossing points at temperature $T_F$ for the polynomial coupling with $\beta_p=125$, together with the jumps at $T_F$.}
\label{tab:poly-crossings}
\setlength{\tabcolsep}{4pt}
\begin{tabular*}{\textwidth}{@{\extracolsep{\fill}}rrrrrr@{}}
\toprule
Coupling & $\epsilon$ & $M_S/\lambda$ & $\lambda T_F$ & $\Delta S_{T_F}/\lambda^2$ & $T_F\Delta S_{T_F}/\lambda$ \\
\midrule
$\zeta_p$ & 1 & 0.0514893 & 0.782674 & \(1.66\times10^{-3}\) & \(1.30\times10^{-3}\) \\
$\zeta_p$ & -1 & 0.0494875 & 0.793491 & \(-1.66\times10^{-3}\) & \(-1.32\times10^{-3}\) \\
\bottomrule
\end{tabular*}
\end{table}

\FloatBarrier

\begin{figure}[htbp]
\centering
\subfigure[Canonical]{\label{fig:exp-free-delta}%
\includegraphics[width=0.485\textwidth]{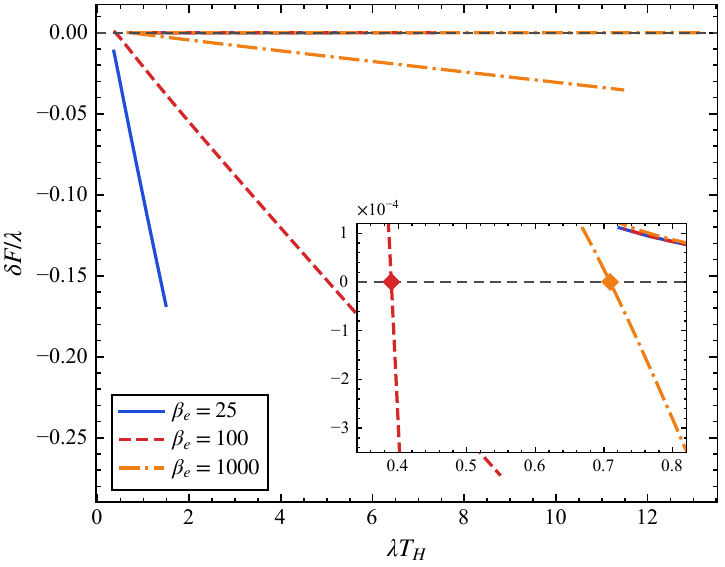}}%
\hfill
\subfigure[Phantom]{\label{fig:exp-free-delta-phantom}%
\includegraphics[width=0.485\textwidth]{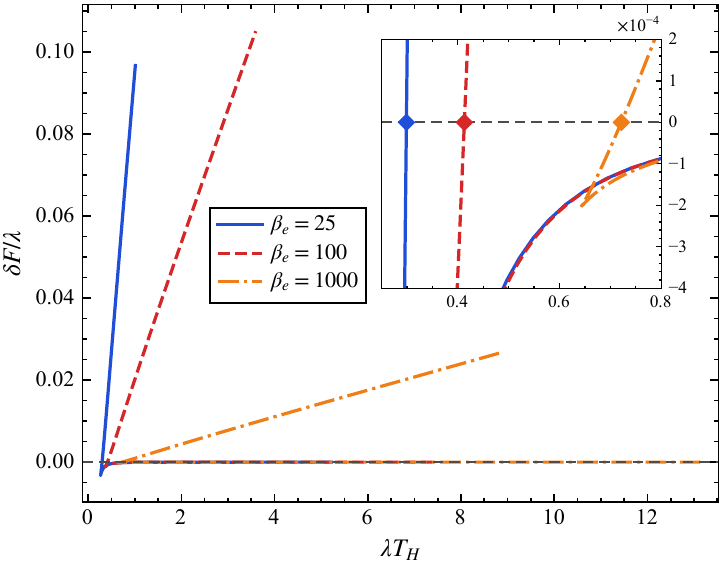}}%
\caption{Free-energy difference from Schwarzschild at the same
temperature for exponential black holes. Line styles distinguish
$\beta_e=25$, $100$ and $1000$. Insets resolve the crossings, marked by
diamonds; the horizontal reference line is $\delta F=0$. Curves are not
connected across unresolved intervals.}
\label{fig:exp-free-phase}
\end{figure}

The exponential families show the same reversal
[Fig.~\ref{fig:exp-free-phase}]. Canonical solutions cross from
$\delta F>0$ to $\delta F<0$ for $\beta_e=100$ and $1000$ as $\phi_H$
increases. For $\beta_e=25$, the larger-$\phi_H$ segment has $\delta F<0$,
with no crossing resolved in the computed intervals.
The phantom lower branches and portions of the primary branches have
$\delta F<0$. On each phantom primary branch, $\delta F$ changes from
negative to positive as $x$ increases through $x_F$.
The phantom free-energy roots are $x_F\simeq1.23714$, $1.24085$ and $1.24288$
for $\beta_e=25$, $100$ and $1000$.
Table~\ref{tab:exp-crossings} lists the entropy crossings and the values
$(T_F,M_F,Q_{s,F})$, together with the jumps.

\begin{table}[htbp]
\centering\small
\caption{Entropy crossing points at mass $M_S$ and free-energy crossing points at $(T_F,M_F,Q_{s,F})$ for the exponential coupling, together with the jumps at $T_F$.}
\label{tab:exp-crossings}
\setlength{\tabcolsep}{4pt}
\begin{tabular*}{\textwidth}{@{\extracolsep{\fill}}rrrrrrrr@{}}
\toprule
$\epsilon$ & $\beta_e$ & $M_S/\lambda$ & $\lambda T_F$ & $M_F/\lambda$ & $Q_{s,F}/\lambda$ & $\Delta S_{T_F}/\lambda^2$ & $T_F\Delta S_{T_F}/\lambda$ \\
\midrule
1 & 100 & 0.107804 & 0.390017 & 0.113141 & 0.034368 & 0.028520 & 0.011123 \\
1 & 1000 & 0.057167 & 0.708881 & 0.058179 & 0.012077 & 0.002892 & 0.002050 \\
-1 & 25 & 0.117251 & 0.299068 & 0.098174 & 0.093589 & $-0.116592$ & $-0.034869$ \\
-1 & 100 & 0.090578 & 0.413011 & 0.084296 & 0.043500 & $-0.029157$ & $-0.012042$ \\
-1 & 1000 & 0.054103 & 0.721610 & 0.053040 & 0.012946 & $-0.002908$ & $-0.002099$ \\
\bottomrule
\end{tabular*}
\end{table}

\FloatBarrier

\begin{figure}[htbp]
\centering
\subfigure[]{\label{fig:exp-differences-full}%
\includegraphics[width=0.485\textwidth,trim=0bp 247.2bp 0bp 0bp,clip]{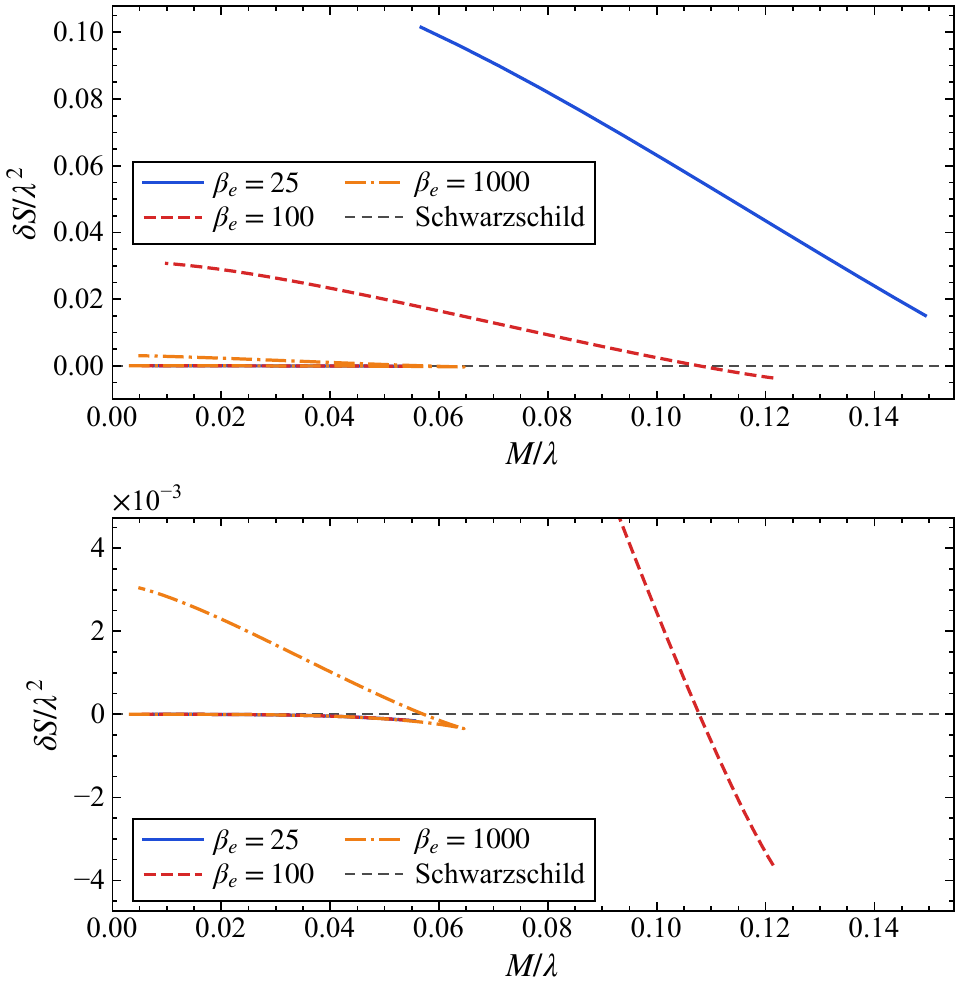}}%
\hfill
\subfigure[]{\label{fig:exp-differences-zoom}%
\includegraphics[width=0.485\textwidth,trim=0bp 0bp 0bp 228bp,clip]{figures/fig14.pdf}}%
\caption{Entropy difference from Schwarzschild at the same mass for
canonical exponential black holes. Panel (a) shows the computed range;
panel (b) enlarges the vertical scale around $\delta S=0$.
Line styles distinguish $\beta_e=25$, $100$ and $1000$. Gaps denote
unresolved numerical intervals.}
\label{fig:exp-differences}
\end{figure}

For the canonical exponential solutions, Fig.~\ref{fig:exp-differences}
resolves the entropy crossings at fixed mass. We also check their
normalization against Ref.~\cite{Herdeiro:2026sur}, using
\begin{equation}
\varphi=2\phi,\qquad \alpha_{\rm ref}=\lambda^2,\qquad
Q_{\rm ref}=2Q_s,\qquad
f_{\rm(iii)}(\varphi)
=\frac{1-e^{-\beta\varphi^4/16}}{4\beta}.
\label{eq:convention-map}
\end{equation}
Then $f_{\rm(iii)}(2\phi)=\zeta_e(\phi)$ for $\beta=\beta_e$.
For $\beta_e=1000$, our canonical crossing temperature
$\lambda T_F\simeq0.708881$ agrees with $0.70894$ reported there.

Table~\ref{tab:coupling-comparison} compares the canonical polynomial and
exponential families whose couplings agree through order $\phi^8$
[Eq.~\eqref{eq:coupling-match}]. Their free-energy crossings occur at
$u\simeq0.95$ and $2.40$, respectively, with $u$ defined in
Eq.~\eqref{eq:expansion-domain}. Both lie beyond the small-field regime
$u\ll1$, explaining why the higher-order coupling terms affect the
crossing quantities and extrema.

\begin{table}[htbp]
\centering\small
\caption{Canonical polynomial and exponential couplings matched through order $\phi^8$. The crossing points and extrema refer to the solution family of each coupling.}
\label{tab:coupling-comparison}
\begin{tabular}{lrr}
\toprule
Quantity & Polynomial $\zeta_p$ & Exponential $\zeta_e$\\
\midrule
Coupling parameter & $\beta_p=125$ & $\beta_e=1000$\\
$M_S/\lambda$ & $0.0514893$ & $0.057167$\\
$\lambda T_F$ & $0.782674$ & $0.708881$\\
$\Delta S_{T_F}/\lambda^2$ & $0.00166$ & $0.002892$\\
Temperature minimum $x_T$ & $0.90289$ & $1.0431$\\
Mass maximum $x_M$ & $0.90700$ & $1.0546$\\
\bottomrule
\end{tabular}
\end{table}

\FloatBarrier

For both couplings, Tables~\ref{tab:poly-crossings} and
\ref{tab:exp-crossings} show positive entropy and mass jumps for
canonical solutions and negative jumps for phantom solutions.
The slope $[\ud(\delta F)/\ud T_H]_{T_F}=-\Delta S_{T_F}$ therefore
gives opposite crossing directions as temperature increases.
The scalar charge distinguishes the two configurations at a crossing:
$Q_s=0$ for Schwarzschild and $Q_{s,F}\ne0$ for the scalarized solution
\cite{Herdeiro:2026sur}.
The finite jumps were interpreted as first-order phase transitions in
canonical EsGB theory \cite{Zou:2026thermo,Herdeiro:2026sur}.
For both canonical and phantom families, the crossing solutions lie
outside the $C>0$ intervals in Sec.~\ref{sec:thermal-response} and have
negative heat capacity, as does the asymptotically flat Schwarzschild
reference \cite{Gibbons:1976ue,Gross:1982cv}.
The free-energy comparison favors the phantom lower branches and the
portions of their primary branches with $\delta F<0$ over Schwarzschild
black holes at the same temperature.

\FloatBarrier

\section{Discussion and conclusions}
\label{sec:comparison}
\label{con}

In this work, we have investigated the thermodynamic properties of static,
nonlinearly scalarized black holes in EPsGB theory with polynomial and
exponential couplings. We have compared the fully backreacted solutions
with the corresponding canonical EsGB black holes
\cite{Zou:2026thermo,Herdeiro:2026sur} and calculated their Hawking
temperature and Wald entropy. We have also checked the first law
numerically for representative solutions.
For both couplings, the phantom lower branches have larger Wald
entropy at fixed mass and lower Helmholtz free energy at fixed
temperature than Schwarzschild black holes, reversing the canonical
ordering. Both advantages extend into the primary branches up to their
respective entropy and free-energy crossings.
The phantom free-energy advantage also persists in the pure quartic
limit, where the canonical solutions remain disfavored relative to
Schwarzschild.
The reversal involves both the explicit Gauss--Bonnet
entropy correction and the change in horizon area due to metric
backreaction. On the phantom exponential primary branches, the
negative Gauss--Bonnet contribution exceeds the area term and gives
negative Wald entropy at small horizon radii.

For the canonical polynomial and exponential couplings matched through
order $\phi^8$, we find that the mass and temperature curves nearly
coincide at small scalar amplitude. At finite amplitude, higher-order coupling
terms modify the nonlinear scalar sources and metric backreaction.
Consequently, the mass and temperature extrema and the free-energy
crossing temperatures differ between the two couplings. The crossings lie beyond the
small-field regime.
Positive heat capacity occurs between the mass maximum and the
temperature minimum. As the horizon scalar increases, the temperature
minimum precedes the mass maximum
for the canonical solutions, while their order is reversed for the
phantom solutions. The pure quartic solutions have negative heat
capacity throughout the range studied.

At the free-energy crossings, the entropy and mass jumps are
positive for canonical black holes and negative for phantom black
holes. The nonzero jumps obey $\Delta M_{T_F}=T_F\Delta S_{T_F}$.
The scalar charge takes the distinct values $Q_s=0$ and $Q_{s,F}\ne0$
for the Schwarzschild and scalarized solutions at the crossing.
The thermodynamic phase structure can also be investigated in a finite
cavity or asymptotically AdS spacetime; scalarized AdS black holes have
already been constructed in EsGB theory \cite{Zou:2022hjj}.
For both polynomial and exponential couplings, the phantom lower
branches and portions of the primary branches are thermodynamically
preferred over Schwarzschild black holes at fixed temperature.
As temperature increases through a primary-branch crossing, canonical
scalarized black holes become preferred, whereas phantom black holes
become disfavored. The phantom field thus reverses both the lower-branch
ordering and the direction of the primary-branch exchange of
thermodynamic preference with Schwarzschild black holes.

\FloatBarrier

\section*{Acknowledgments}

This research is supported by the Natural Science Basic Research Program
of Shaanxi Province (Grant No.~2026JC-YBMS-0083), the Jiangxi Provincial
Natural Science Foundation (Grant No.~20262BAC240347), and the National
Natural Science Foundation of China (NNSFC) (Grant No.~12365009).

\begingroup
\small
\bibliographystyle{unsrt}
\bibliography{references}

@article{Damour:1993hw,
  author = {T.~Damour and G.~Esposito-Far\`ese},
  title = {{Nonperturbative strong-field effects in tensor-scalar theories of gravitation}},
  journal = {Phys. Rev. Lett.},
  volume = {70},
  pages = {2220},
  year = {1993},
}

@article{Bekenstein:1995un,
  author = {J.~D.~Bekenstein},
  title = {{Novel no-scalar-hair theorem for black holes}},
  journal = {Phys. Rev. D},
  volume = {51},
  pages = {R6608},
  year = {1995},
}

@article{Herdeiro:2015waa,
  author = {C.~A.~R.~Herdeiro and E.~Radu},
  title = {{Asymptotically flat black holes with scalar hair: a review}},
  journal = {Int. J. Mod. Phys. D},
  volume = {24},
  pages = {1542014},
  year = {2015},
  eprint = {1504.08209},
  archiveprefix = {arXiv},
  primaryclass = {gr-qc},
  note = {[arXiv:1504.08209 [gr-qc]]},
}

@article{Kanti:1995vq,
  author = {P.~Kanti and N.~E.~Mavromatos and J.~Rizos and K.~Tamvakis and E.~Winstanley},
  title = {{Dilatonic black holes in higher curvature string gravity}},
  journal = {Phys. Rev. D},
  volume = {54},
  pages = {5049--5058},
  year = {1996},
  eprint = {hep-th/9511071},
  archiveprefix = {arXiv},
  note = {[arXiv:hep-th/9511071]},
}

@article{Sotiriou:2013qea,
  author = {T.~P.~Sotiriou and S.~Y.~Zhou},
  title = {{Black hole hair in generalized scalar-tensor gravity}},
  journal = {Phys. Rev. Lett.},
  volume = {112},
  pages = {251102},
  year = {2014},
  eprint = {1312.3622},
  archiveprefix = {arXiv},
  primaryclass = {gr-qc},
  note = {[arXiv:1312.3622 [gr-qc]]},
}

@article{Doneva:2017bvd,
  author = {D.~D.~Doneva and S.~S.~Yazadjiev},
  title = {{New Gauss-Bonnet black holes with curvature-induced scalarization in extended scalar-tensor theories}},
  journal = {Phys. Rev. Lett.},
  volume = {120},
  pages = {131103},
  year = {2018},
  eprint = {1711.01187},
  archiveprefix = {arXiv},
  primaryclass = {gr-qc},
  note = {[arXiv:1711.01187 [gr-qc]]},
}

@article{Silva:2017uqg,
  author = {H.~O.~Silva and J.~Sakstein and L.~Gualtieri and T.~P.~Sotiriou and E.~Berti},
  title = {{Spontaneous scalarization of black holes and compact stars from a Gauss-Bonnet coupling}},
  journal = {Phys. Rev. Lett.},
  volume = {120},
  pages = {131104},
  year = {2018},
  eprint = {1711.02080},
  archiveprefix = {arXiv},
  primaryclass = {gr-qc},
  note = {[arXiv:1711.02080 [gr-qc]]},
}

@article{Antoniou:2017acq,
  author = {G.~Antoniou and A.~Bakopoulos and P.~Kanti},
  title = {{Evasion of no-hair theorems and novel black-hole solutions in Gauss-Bonnet theories}},
  journal = {Phys. Rev. Lett.},
  volume = {120},
  pages = {131102},
  year = {2018},
  eprint = {1711.03390},
  archiveprefix = {arXiv},
  primaryclass = {hep-th},
  note = {[arXiv:1711.03390 [hep-th]]},
}

@article{Antoniou:2017hxj,
  author = {G.~Antoniou and A.~Bakopoulos and P.~Kanti},
  title = {{Black-hole solutions with scalar hair in Einstein-scalar-Gauss-Bonnet theories}},
  journal = {Phys. Rev. D},
  volume = {97},
  pages = {084037},
  year = {2018},
  eprint = {1711.07431},
  archiveprefix = {arXiv},
  primaryclass = {hep-th},
  note = {[arXiv:1711.07431 [hep-th]]},
}

@article{Doneva:2022ewd,
  author = {D.~D.~Doneva and F.~M.~Ramazano\u{g}lu and H.~O.~Silva and T.~P.~Sotiriou and S.~S.~Yazadjiev},
  title = {{Spontaneous scalarization}},
  journal = {Rev. Mod. Phys.},
  volume = {96},
  pages = {015004},
  year = {2024},
  eprint = {2211.01766},
  archiveprefix = {arXiv},
  primaryclass = {gr-qc},
  note = {[arXiv:2211.01766 [gr-qc]]},
}

@article{Blazquez-Salcedo:2018jnn,
  author = {J.~L.~Bl\'azquez-Salcedo and D.~D.~Doneva and J.~Kunz and S.~S.~Yazadjiev},
  title = {{Radial perturbations of the scalarized Einstein-Gauss-Bonnet black holes}},
  journal = {Phys. Rev. D},
  volume = {98},
  pages = {084011},
  year = {2018},
  eprint = {1805.05755},
  archiveprefix = {arXiv},
  primaryclass = {gr-qc},
  note = {[arXiv:1805.05755 [gr-qc]]},
}

@article{Silva:2018qhn,
  author = {H.~O.~Silva and C.~F.~B.~Macedo and T.~P.~Sotiriou and L.~Gualtieri and J.~Sakstein and E.~Berti},
  title = {{Stability of scalarized black hole solutions in scalar-Gauss-Bonnet gravity}},
  journal = {Phys. Rev. D},
  volume = {99},
  pages = {064011},
  year = {2019},
  eprint = {1812.05590},
  archiveprefix = {arXiv},
  primaryclass = {gr-qc},
  note = {[arXiv:1812.05590 [gr-qc]]},
}

@article{Blazquez-Salcedo:2020rhf,
  author = {J.~L.~Bl\'azquez-Salcedo and D.~D.~Doneva and S.~Kahlen and J.~Kunz and P.~Nedkova and S.~S.~Yazadjiev},
  title = {{Axial perturbations of the scalarized Einstein-Gauss-Bonnet black holes}},
  journal = {Phys. Rev. D},
  volume = {101},
  pages = {104006},
  year = {2020},
  eprint = {2003.02862},
  archiveprefix = {arXiv},
  primaryclass = {gr-qc},
  note = {[arXiv:2003.02862 [gr-qc]]},
}

@article{Blazquez-Salcedo:2020caw,
  author = {J.~L.~Bl\'azquez-Salcedo and D.~D.~Doneva and S.~Kahlen and J.~Kunz and P.~Nedkova and S.~S.~Yazadjiev},
  title = {{Polar quasinormal modes of the scalarized Einstein-Gauss-Bonnet black holes}},
  journal = {Phys. Rev. D},
  volume = {102},
  pages = {024086},
  year = {2020},
  eprint = {2006.06006},
  archiveprefix = {arXiv},
  primaryclass = {gr-qc},
  note = {[arXiv:2006.06006 [gr-qc]]},
}

@article{Minamitsuji:2024twp,
  author = {M.~Minamitsuji and S.~Mukohyama and S.~Tsujikawa},
  title = {{Angular and radial stabilities of spontaneously scalarized black holes in the presence of scalar-Gauss-Bonnet couplings}},
  journal = {Phys. Rev. D},
  volume = {109},
  pages = {104057},
  year = {2024},
  eprint = {2403.10048},
  archiveprefix = {arXiv},
  primaryclass = {gr-qc},
  note = {[arXiv:2403.10048 [gr-qc]]},
}

@article{Macedo:2019sem,
  author = {C.~F.~B.~Macedo and J.~Sakstein and E.~Berti and L.~Gualtieri and H.~O.~Silva and T.~P.~Sotiriou},
  title = {{Self-interactions and spontaneous black hole scalarization}},
  journal = {Phys. Rev. D},
  volume = {99},
  pages = {104041},
  year = {2019},
  eprint = {1903.06784},
  archiveprefix = {arXiv},
  primaryclass = {gr-qc},
  note = {[arXiv:1903.06784 [gr-qc]]},
}

@article{Doneva:2022yqu,
  author = {D.~D.~Doneva and L.~G.~Collodel and S.~S.~Yazadjiev},
  title = {{Spontaneous nonlinear scalarization of Kerr black holes}},
  journal = {Phys. Rev. D},
  volume = {106},
  pages = {104027},
  year = {2022},
  eprint = {2208.02077},
  archiveprefix = {arXiv},
  primaryclass = {gr-qc},
  note = {[arXiv:2208.02077 [gr-qc]]},
}

@article{Lai:2023gwe,
  author = {M.~Y.~Lai and D.~C.~Zou and R.~H.~Yue and Y.~S.~Myung},
  title = {{Nonlinearly scalarized rotating black holes in Einstein-scalar-Gauss-Bonnet theory}},
  journal = {Phys. Rev. D},
  volume = {108},
  pages = {084007},
  year = {2023},
  eprint = {2304.08012},
  archiveprefix = {arXiv},
  primaryclass = {gr-qc},
  note = {[arXiv:2304.08012 [gr-qc]]},
}

@article{Doneva:2021tvn,
  author = {D.~D.~Doneva and S.~S.~Yazadjiev},
  title = {{Beyond the spontaneous scalarization: New fully nonlinear mechanism for the formation of scalarized black holes and its dynamical development}},
  journal = {Phys. Rev. D},
  volume = {105},
  pages = {L041502},
  year = {2022},
  eprint = {2107.01738},
  archiveprefix = {arXiv},
  primaryclass = {gr-qc},
  note = {[arXiv:2107.01738 [gr-qc]]},
}

@article{Blazquez-Salcedo:2022omw,
  author = {J.~L.~Bl\'azquez-Salcedo and D.~D.~Doneva and J.~Kunz and S.~S.~Yazadjiev},
  title = {{Radial perturbations of scalar-Gauss-Bonnet black holes beyond spontaneous scalarization}},
  journal = {Phys. Rev. D},
  volume = {105},
  pages = {124005},
  year = {2022},
  eprint = {2203.00709},
  archiveprefix = {arXiv},
  primaryclass = {gr-qc},
  note = {[arXiv:2203.00709 [gr-qc]]},
}

@article{Pombo:2023lxg,
  author = {A.~M.~Pombo and D.~D.~Doneva},
  title = {{Effects of mass and self-interaction on nonlinear scalarization of scalar-Gauss-Bonnet black holes}},
  journal = {Phys. Rev. D},
  volume = {108},
  pages = {124068},
  year = {2023},
  eprint = {2310.08638},
  archiveprefix = {arXiv},
  primaryclass = {gr-qc},
  note = {[arXiv:2310.08638 [gr-qc]]},
}

@article{Zou:2024paper,
  author = {D.~C.~Zou and X.~Yang and M.~Y.~Lai and H.~Huang and B.~Liu and J.~Kunz and Y.~S.~Myung and R.~H.~Yue},
  title = {{Existence of nonlinearly scalarized black holes in Einstein-scalar-Gauss-Bonnet theory with polynomial couplings}},
  journal = {Phys. Rev. D},
  volume = {113},
  pages = {104015},
  year = {2026},
  doi = {10.1103/y3sr-wd4q},
  eprint = {2404.19521},
  archiveprefix = {arXiv},
  primaryclass = {gr-qc},
  note = {doi:10.1103/y3sr-wd4q, [arXiv:2404.19521 [gr-qc]]},
}

@article{Wald:1993nt,
  author = {R.~M.~Wald},
  title = {{Black hole entropy is the Noether charge}},
  journal = {Phys. Rev. D},
  volume = {48},
  pages = {R3427},
  year = {1993},
  eprint = {gr-qc/9307038},
  archiveprefix = {arXiv},
  note = {[arXiv:gr-qc/9307038]},
}

@article{Iyer:1994ys,
  author = {V.~Iyer and R.~M.~Wald},
  title = {{Some properties of Noether charge and a proposal for dynamical black hole entropy}},
  journal = {Phys. Rev. D},
  volume = {50},
  pages = {846},
  year = {1994},
  eprint = {gr-qc/9403028},
  archiveprefix = {arXiv},
  note = {[arXiv:gr-qc/9403028]},
}

@article{Zou:2026thermo,
  author = {D.~C.~Zou and X.~Yang and M.~Y.~Lai and H.~Huang and Y.~S.~Myung},
  title = {{Thermodynamics and phase transitions of nonlinearly scalarized black holes in Einstein-scalar-Gauss-Bonnet theory}},
  journal = {Phys. Lett. B},
  volume = {878},
  pages = {140573},
  year = {2026},
  eprint = {2604.20153},
  archiveprefix = {arXiv},
  primaryclass = {gr-qc},
  note = {[arXiv:2604.20153 [gr-qc]]},
}

@article{Li:2025vcq,
    author = {Z.~Li and H.~S.~Liu and H.~L\"u},
    title = "{From wormhole to black hole with negative mass}",
    eprint = "2509.20755",
    archivePrefix = "arXiv",
    primaryClass = "gr-qc",
    doi = "10.1103/9h7h-686p",
    journal = "Phys. Rev. D",
    volume = "113",
    number = "4",
    pages = "044032",
    year = "2026"
}

@article{Herdeiro:2026sur,
  author = {C.~Herdeiro and H.~Huang and J.~Kunz and M.~Y.~Lai and E.~Radu and D.~C.~Zou},
  title = {{Phase structure of scalarized black holes in Einstein-scalar-Gauss-Bonnet gravity}},
  journal = {Phys. Rev. D},
  volume = {113},
  pages = {124046},
  year = {2026},
  doi = {10.1103/4k1l-jzvz},
  eprint = {2603.24164},
  archiveprefix = {arXiv},
  primaryclass = {gr-qc},
  note = {[arXiv:2603.24164 [gr-qc]]},
}

@article{Gibbons:1976ue,
  author = {G.~W.~Gibbons and S.~W.~Hawking},
  title = {{Action integrals and partition functions in quantum gravity}},
  journal = {Phys. Rev. D},
  volume = {15},
  pages = {2752},
  year = {1977},
}

@article{Gross:1982cv,
  author = {D.~J.~Gross and M.~J.~Perry and L.~G.~Yaffe},
  title = {{Instability of flat space at finite temperature}},
  journal = {Phys. Rev. D},
  volume = {25},
  pages = {330},
  year = {1982},
}

@article{Bekenstein:1973ur,
  author = {Jacob D. Bekenstein},
  title = {{Black Holes and Entropy}},
  journal = {Phys. Rev. D},
  volume = {7},
  pages = {2333--2346},
  year = {1973},
  doi = {10.1103/physrevd.7.2333},
}

@article{Hawking:1974sw,
  author = {S. W. Hawking},
  title = {{Particle creation by black holes}},
  journal = {Commun. Math. Phys.},
  volume = {43},
  pages = {199--220},
  year = {1975},
  doi = {10.1007/bf02345020},
}

@article{York:1986it,
  author = {James W. York},
  title = {{Black-hole thermodynamics and the Euclidean Einstein action}},
  journal = {Phys. Rev. D},
  volume = {33},
  pages = {2092--2099},
  year = {1986},
  doi = {10.1103/physrevd.33.2092},
}

@article{Lai:2026vxs,
  author = {M.~Y.~Lai and H.~Huang and J.~Kunz and Y.~S.~Myung and D.~C.~Zou},
  title = {{Spin-induced nonlinear scalarization of Kerr black holes in Einstein-scalar-Gauss-Bonnet gravity}},
  journal = {Phys. Rev. D},
  volume = {114},
  pages = {024007},
  year = {2026},
  doi = {10.1103/tp1r-hhq6},
  eprint = {2604.27811},
  archiveprefix = {arXiv},
  primaryclass = {gr-qc},
  note = {[arXiv:2604.27811 [gr-qc]]},
}

@article{Zou:2022hjj,
  author = {D.~C.~Zou and B.~Meng and M.~Zhang and S.~Y.~Li and M.~Y.~Lai and Y.~S.~Myung},
  title = {{Analytical Approximate Solutions for Scalarized AdS Black Holes}},
  journal = {Universe},
  volume = {9},
  pages = {26},
  year = {2023},
  doi = {10.3390/universe9010026},
  eprint = {2301.04784},
  archiveprefix = {arXiv},
  primaryclass = {gr-qc},
  note = {[arXiv:2301.04784 [gr-qc]]},
}
\endgroup
\end{document}